\documentclass[10pt]{article}

\usepackage{arxiv}
\usepackage[T1]{fontenc}
\usepackage[utf8x]{inputenc}
\usepackage[table,dvipsnames]{xcolor} 

\usepackage{hyperref,url,multirow,amsfonts,amsmath,amssymb,nicefrac,graphicx}
\usepackage{longtable}
\usepackage{orcidlink} 
\usepackage{epigraph}

\definecolor{fill}{rgb}{0.862,0.886,0.917} 

\newcommand*\rot{\rotatebox{90}}
\newcommand*\ch{$\bullet$}
\newcommand*\me{$\bigcirc$}
\newcommand*\eg{$\odot$}
\newcommand*\na{$\blacktriangledown$}
\newcommand*\sa{$\blacktriangle$}
\newcommand*\po{$\square$}
\newcommand*\an{$\triangledown$}
\newcommand*\as{$\vartriangle$}
\newcommand*\india{$\blacksquare$}
\newcommand*\ms{$\blacklozenge$}

\title{Parallels in the symbolism of star constellations} 

\date{} 
\author{
	Doina Bucur \orcidlink{0000-0002-4830-7162}\\
	\href{mailto:d.bucur@utwente.nl}{\nolinkurl{d.bucur@utwente.nl}}\\
	University of Twente, The Netherlands
}

\begin{document}
\maketitle


\begin{abstract}
We answer the question whether, when forming constellations in the night sky, people in astronomical cultures around the world and through time consistently imagined and assigned the same symbolism to the same (type of) star group. Evidence of semantic parallels has so far been anecdotal. We use two complementary definitions for a star group: (1) a star group in a fixed region of the sky (regardless of its exact star composition), and (2) a star group with a particular shape and brightness (regardless of its location in the sky). Over a dataset of 2003 constellations from 82 astronomical cultures, we find many semantic parallels which are likely naturally induced by the shape and composition of the star pattern. In certain cultural regions, geometric and group symbols are perceived consistently over small and uniformly bright star groups, naturalistic humanoids in large star groups with non-linear minimum spanning tree (MST) and stars inside the convex hull, and reptiles in star groups with low aspect ratio or linear MST. These naturally induced semantics, seemingly endogenous to certain sky patterns, show that there are universal (rather than learnt) patterns behind forming and naming constellations.

\end{abstract}




\section{Introduction}



Constellation figures are geometric representations at the intersection of nature and culture. The stars in a constellation can be chosen and a meaning freely assigned (a bear, a scorpion), but that same constellation is constrained to a background of stars shared by all astronomical cultures around the world. We ask where and to what extent \textbf{constellation semantics} are \textbf{universal}. Semantic parallels may be caused by the star pattern itself, which naturally resembles the entity named; this is an natural (or endogenous) effect of the star pattern. It may also be caused by a common cultural influence, which imposes one common view upon a sky region; this is the cultural (or exogenous) effect of the human imagination. We quantify semantic universality, and, when evidence supports it, point to a cause (natural or cultural).

A \textbf{natural effect} of the star pattern on constellation semantics was not measured systematically, but has anecdotal evidence, as follows. The dipper symbolism for Big Dipper appears from the West to China and rarely in N America, although it remains rare~\cite{berezkin2012seven}---it was thus ascribed to the star pattern naturally resembling this man-made tool~\cite{gibbon1964asiatic}. There are parallels between Western and indigenous American constellations to an extent unlikely to be due to colonial influence: constellations located in the sky region defined by the International Astronomical Union~\cite{IAU} ({\bf IAU}) as Sco, with scorpion geometry and semantics, were documented also in the pre-colonial Aztec and Maya cultures of Mesoamerica~\cite{kelley2005exploring}. A small-scale cognitive study showed that the line geometry for 30 of the classical Ptolemaic constellations~\cite{toomer1984ptolemy} are predictable to Western observers from the star pattern alone~\cite{dry2009perceptual}. Also, worldwide, constellations adjacent to 35\% of popular stars have rather universal line geometry (these are located in IAU Sco, CrB, Cas, and to a lesser extent UMa and Leo)~\cite{bucur2022pone}, but it is not known \emph{why}, nor whether their symbolism is also universal.

\textbf{Cultural diffusion} has been observed. The star groups and their names have retained similarities across astronomical cultures with a common \emph{ancestry}, even when distant in time or space. This may be due to variables associated with cultural phylogeny: \emph{geolocation} (cultures of the tropics form different astronomical systems than those in temperate zones~\cite{aveni1981tropical}), \emph{astronomical literacy} (in the region influenced by ancient China, detailed star charts were maintained, and also faint stars were linked into constellations~\cite{sun1997chinese}), and \emph{cultural myths or practices}. The latter likely have a large influence upon constellation design. For example, the cart symbolism for Big Dipper is known to all Indo-European traditions in Europe (which use carts) and not to others outside their cultural region~\cite{berezkin2012seven}. In cultures with Mesopotamian or Mediterranean origin, many constellations represent human and animal mythological characters~\cite{rogers1998origins1,rogers1998origins2}, while Northern Dene tribes in Alaska and Canada universally draw a whole-sky Traveler constellation of their principal mythological character~\cite{cannon2021northern}. A bird in flight dominates the skies of Polynesian cultures of the south Pacific~\cite{chadwick2017great}, and there are parallels in symbolism between pre-Columbian N-American and central and west Asian myth for the Big Dipper~\cite{gibbon1964asiatic} and Orion's belt asterisms~\cite{gibbon1972asiatic}, due to a distant common origin.


\begin{figure}[!htb]
	\centering
	\includegraphics[width=\linewidth]{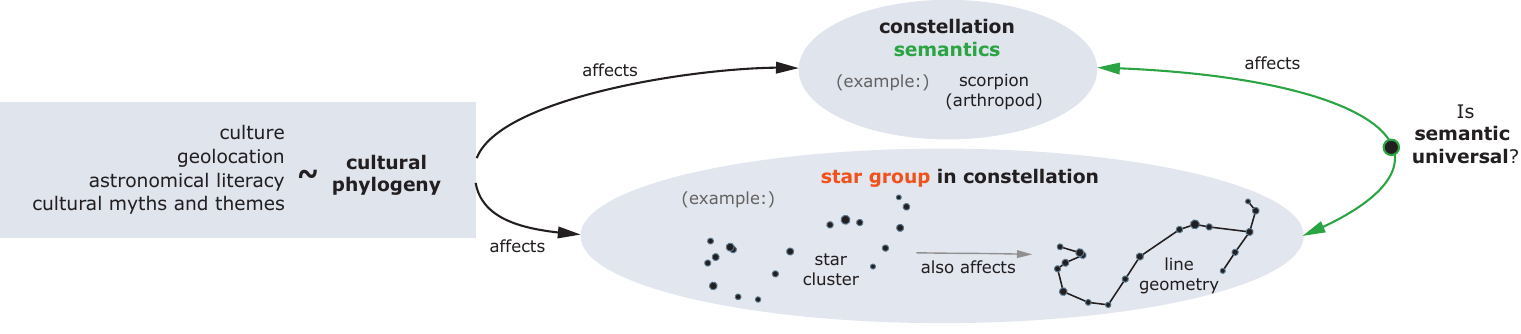}
	\caption{{\footnotesize {\bf Causal graph.} Causal effects between culture, constellation semantics, and star group. We answer the question the question \emph{Is a semantic universal?} Culture metadata (phylogeny, etc.) is a confounder.}}
	\label{fig:causal_model}
\end{figure}

We draw a \textbf{causal model} of potential influence upon constellation semantics in Figure \ref{fig:causal_model}. All variables can be measured. We answer the question \emph{Is semantic universal?}. An association between a chosen group of stars and semantics would show if and which symbolisms are associated with certain star patterns or sky regions, and are not explained by a common ancestry. Cultural phylogeny is a common cause: cultural myths and themes affect semantics; geolocation and astronomical literacy affect the choice of stars. Between semantics and star group, causal effects may be bidirectional---it is not possible to find which came first in the mind of the sky observer: the star group (perhaps naturally separable in the night sky from other star groups) or the meaning; the meaning or the line geometry. Constellation semantics is the outcome variable; we study potential effects on this. The statistical association between a variable of interest and the outcome can be measured after controlling for cultural phylogeny.

The remaining effect in Figure \ref{fig:causal_model} (between cultural phylogeny and star group or line geometry) is unrelated to semantics, and has been studied. Star group may be (to an extent) natural effects of the human perception of point and brightness patterns~\cite{kemp2020perceptual}, and cultures produce (to an extent) their `signature' star group and line geometries in similar ways: oral astronomies have widespread similarities and use brighter stars across continents, Chinese and Mesopotamian ancestries have opposite geometries, with Polynesian ancestry the only `bridge'~\cite{bucur2022pone}.


\begin{figure}[!htb]
	\centering
	\includegraphics[width=\linewidth]{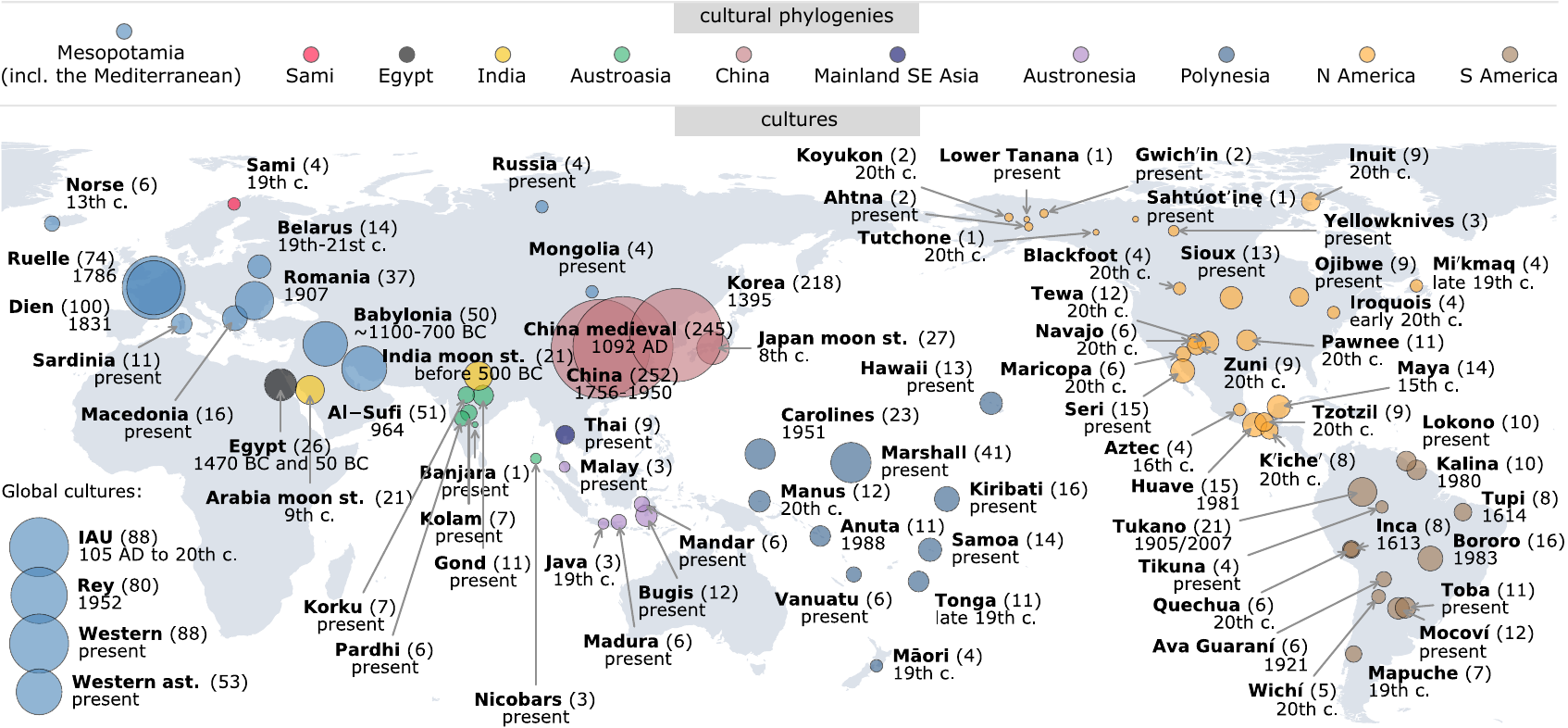}
	\caption{{\footnotesize {\bf Astronomical cultures in the data.} The format is: culture name, (the number of constellations in parentheses), and the documentation date. The area of the data point is proportional to the number of constellations, and the colour follows phylogeny. Cultures with global reach are at the bottom left. The dataset contains 82 cultures and 2003 constellations.}}
	\label{fig:culture_map}
\end{figure}

To answer the question of semantic universality, we use 82 astronomical cultures (mapped in Figure \ref{fig:culture_map}) for which 2003 constellations were identified in terms of star makeup in existing ethnographic, anthropological, or (archeo)astronomical literature. The dataset is open (\url{https://github.com/doinab/constellation-lines}) and was partly studied in~\cite{bucur2022pone}. Included are only constellations or asterisms with \emph{at least two stars} (identified with some certainty) and excluding tight star clusters such as the Pleiades, whose internal pattern of stars is difficult to discern for observers. The cultures are heterogeneous in the number of constellations (between 1 and 252), and the constellations are heterogeneous in number of stars (between 2 and 67) and angular diameter in the sky to an observer on the ground (between 0.24 and 147.74 degrees). Small oral cultures are as interesting to study as large ones: while literate astronomies (Chinese, Egyptian, Mesopotamian/Mediterranean traditions) preserved a whole-sky record, oral astronomies (with vanishing traditions, in the Americas and the Pacific) have a low number of preserved constellations, yet some of these are whole-sky designs.

The {\bf phylogeny} marks a \emph{region of cultural influence or migration}. We have 11 phylogenies. All Western cultures are marked as having {\bf Mesopotamian} ancestry, which oversimplifies a mixed origin: Mesopotamia is the oldest source (of the zodiacal signs), but other Western constellations are Mediterranean, and some minor ones are recent European additions. {\bf Egypt} and {\bf India} denote single ancient astronomical cultures without known external influence. {\bf China} denotes not only cultures located in China itself, but also those in its sphere of influence. {\bf N} and {\bf S America}, {\bf Polynesia}, {\bf Austronesia}, and {\bf Austroasia} denote sets of cultures from migration families, so with cultural similarities. For more information about the data, the smaller phylogenies, and the {\bf semantic annotation}, see Sections~\ref{subsec:data}--\ref{subsec:sem}. 


\section{Results}


We ask if and where an association exists (beyond the scope of single cultural regions) between constellation \emph{star group} and \emph{semantics}. A positive association is a semantic parallel: a star group is assigned the same symbolism.

We cannot use the strong definition of star group (the exact set of stars), because (1) this set is rarely identical across cultures (only the small and bright Big Dipper, Orion's belt, and IAU Cru are commonly recognised), and (2) in some cultures, the exact identity of major or minor stars in the constellation is either not important~\cite{stephenson1994chinese} or uncertain, and can even vary between villages~\cite{vahia2013gond}. Instead, we use two flexible, complementary definitions of a star group:

\begin{description}
	\item[(S.1)] The star group is a \emph{sky region}: a group of stars which includes a fixed ``root star'' (e.g., $\alpha$ CMa in Bayer designation). Star groups over a root star may not be identical across cultures, but always overlap. The root stars are selected based on popularity: they are part of many constellations, and tend to be bright. This definition \emph{fixes} the sky region of the star group on the celestial sphere, but allows the star pattern to vary.
	\item[(S.2)] The star group is a \emph{star pattern}: the number of stars, aspect ratio, spatial diameterin the sky, brightness statistics, and properties of the convex hull and the minimum spanning tree drawn over the star group. Two star groups which are similar in these features need not coincide, and may be located far apart in the sky. This definition \emph{fixes} the features of the star pattern, but allows its location on the celestial sphere to vary.
\end{description}

\subsection{Question (S.1): Semantic universality per sky region}

Significant semantic parallels exist for certain sky regions, outside the bounds of a single cultural region. A semantic parallel may be either natural or cultural: if the cultural regions where it is found may have had exchanges, the primary cause is likely cultural; otherwise, it may be natural. We stratify the analysis by cultural phylogeny, since it is a confounder: it affects both semantics and the choice of stars. (For how phylogenies were determined, see Section~\ref{subsec:data}; for the semantic annotation, Section~\ref{subsec:sem}.)

\begin{figure}[!htb]
	\includegraphics[scale=0.6]{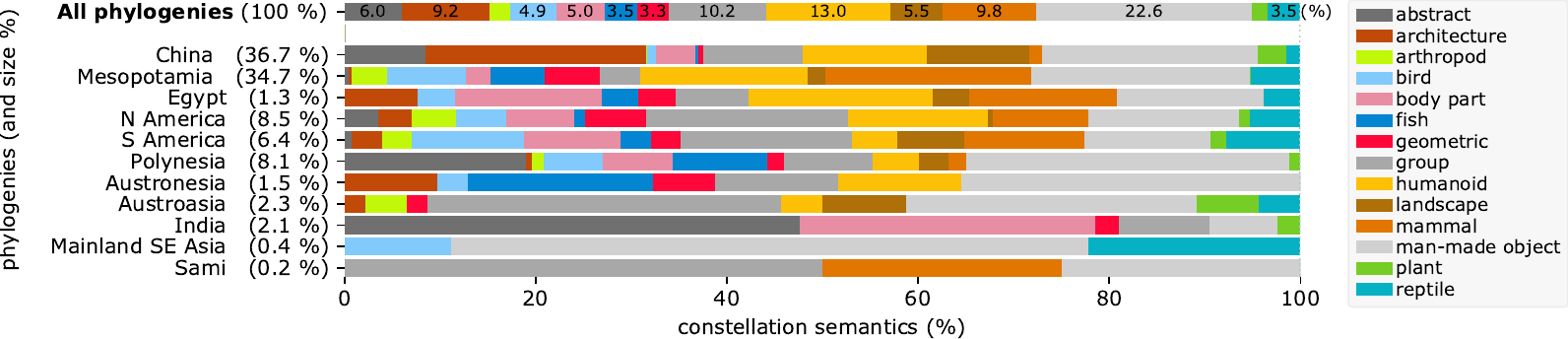}\\[2mm]
	\includegraphics[scale=0.6]{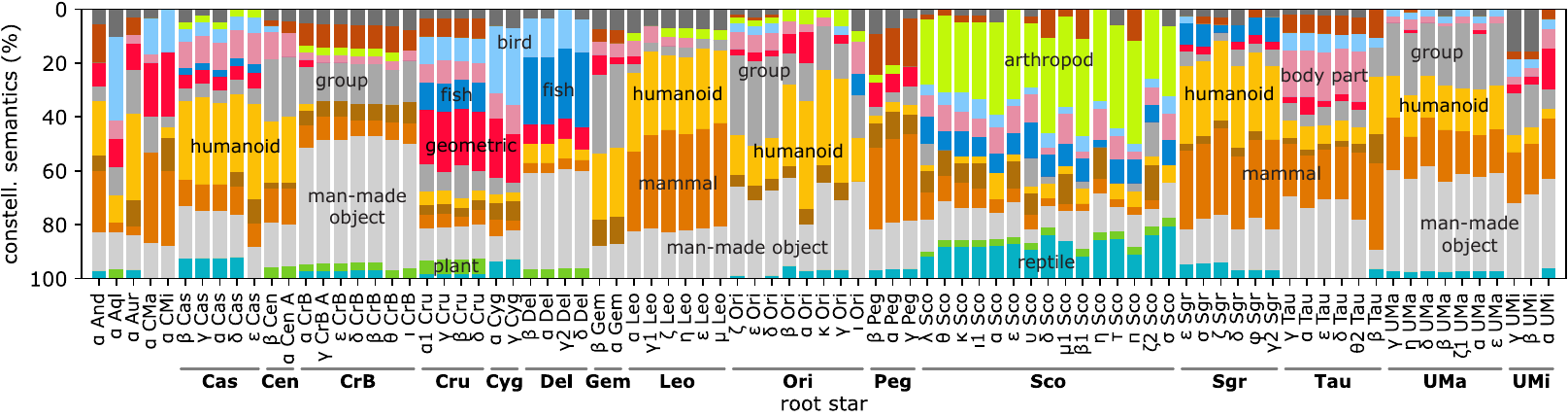}

	\caption{{\small{\bf Constellation semantics.} {\bf (top)} The semantics of 99.45 \% of constellations are identified. This shows their breakdown. (Alternative semantics are all included in the counts.) {\bf (bottom)} Per popular root star worldwide, the breakdown in semantics. Each root star is part of at least 25 (and up to 89) constellations.}}
	\label{fig:breakdown_semantics_phylogeny}
\end{figure}

The cultural regions differ in their semantic makeup (by the semantic breakdown in Figure~\ref{fig:breakdown_semantics_phylogeny}, top). There are dominant phylogenies which may skew the perception of certain semantics: for example, nearly all mammals are in W Asian, Western, and American cultures; they are almost absent from SE Asia and oceanic cultures. Thus, the global semantic counts per root star (Figure~\ref{fig:breakdown_semantics_phylogeny}, bottom) are biased towards large phylogenies, rather than evidence of universal semantics. Typical Western symbolisms appear frequent: the bird for star regions in IAU Aql and Cyg, the arthropod for Sco, the geometric cross for Cru, mammals for CMa, CMi, Leo, Peg, and Tau, many humanoids, and man-made objects for CrB (a crown), UMa and UMi (dippers and carts). Non-Western semantics (fish for Cru, geometric figures for CMa and CMi, body parts for Tau, humanoids in Leo and UMa, and widespread man-made objects) are less frequent. Note that while the IAU sky regions are variable in their number of popular stars (1 in IAU CMa, but 15 in Sco), the semantic breakdown is usually consistent among the root stars, even for large regions such as Sco.

\begin{figure}[!htb]
	\centering

	\includegraphics[scale=0.6]{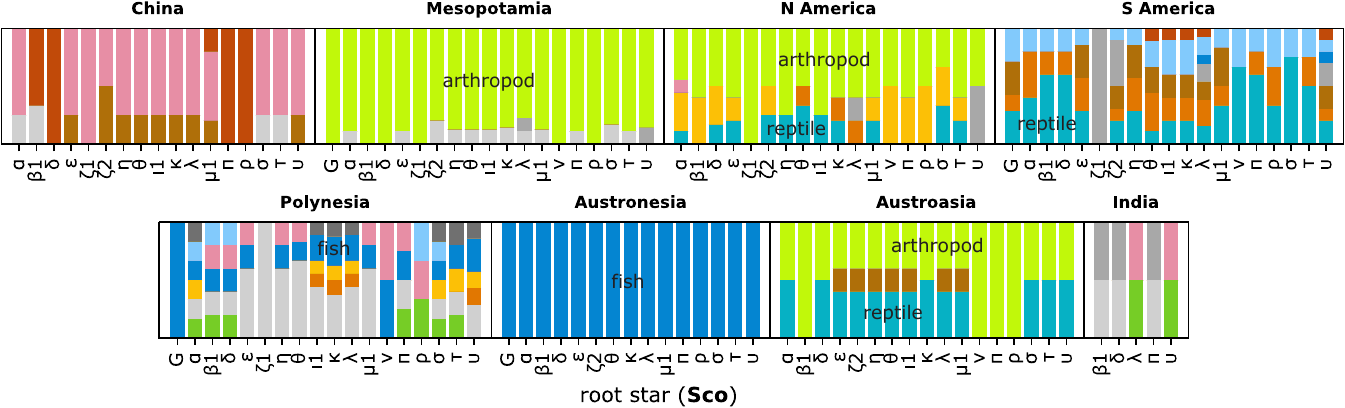}\\[1mm]
	\includegraphics[scale=0.6]{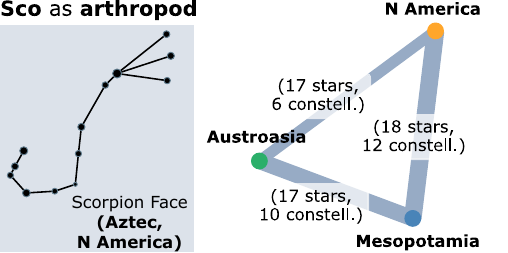}\hspace{10mm}\includegraphics[scale=0.6]{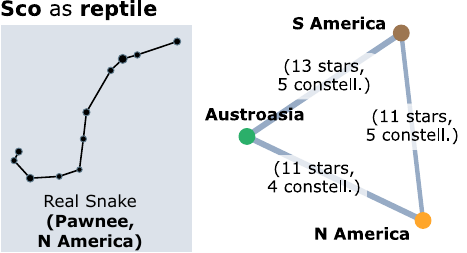}\hspace{10mm}\includegraphics[scale=0.6]{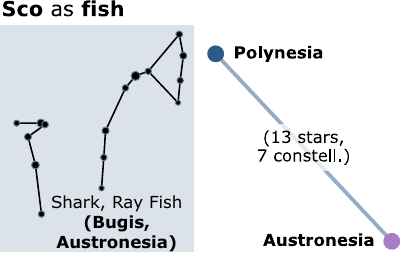}

	\caption{{\small{\bf Semantic parallels for IAU Sco.} {\bf (top)} Per root star and per cultural phylogeny, the breakdown in semantics (semantic colour coding as in Figure~\ref{fig:breakdown_semantics_phylogeny}). Includes root stars with (a) 2+ constellations per phylogeny, and (b) 10+ constellations in the global data. {\bf (bottom)} Three semantic parallels (geometric, reptile, and fish) for the star region of Sco, each in the phylogenies shown by the three similarity networks. The edges are weighted by the similarity score and annotated with the number of stars and constellations in common. Examples are also shown.}}
	\label{fig:breakdown_semantics_Sco}
\end{figure}

Semantic counts stratified by phylogeny provide a more nuanced story. Take the example of the scorpion (arthropod) symbolism for \textbf{IAU Sco}. This symbolism recurs, but only in cultures from three ancestries (Mesopotamian, N American, and Austroasian), and is entirely absent from the others---where different semantic parallels (fish and reptile) exist instead. Figure \ref{fig:breakdown_semantics_Sco} (top) shows the breakdown in semantics of all popular stars in IAU Sco, per phylogeny.

To quantify semantic parallels, we report a \emph{similarity score} between two given phylogenies: averaged over root stars, the joint probability of a semantic occurring in constellations of the same root star. To disregard parallels due to chance, this is normalised by the joint probability of that semantic being present in those phylogenies. Only values greater than 1 are reported (and higher values are more significant). The similarity scores link phylogenies into weighted undirected \emph{similarity networks}, as in Figure \ref{fig:breakdown_semantics_Sco} (bottom); the edge is weighted by the similarity score. Also reported are two additional counts: the \emph{number of stars} and the \emph{number of constellations} in common for that semantic between the two phylogenies (both more significant when larger). (Section~\ref{sec:s1method} provides detail on the method.)

Three semantics for Sco stars have parallels, across different parts of the world (Figure \ref{fig:breakdown_semantics_Sco}, bottom):
\begin{description}
	\item[arthropod] Eight Western, three Mesoamerican (Aztec, Huave, and Tzotzil), and two Austroasian (Gond and Kolam) cultures draw a scorpion figure similar to IAU Sco; mostly the lining of the head varies. An additional Mesoamerican culture (Maya) draws a much smaller scorpion, adjacent to only four Sco stars. 
	\item[reptile] Two N American (Maya, Pawnee), one S American (Tukano), and two Austroasian (Pardhi, Kolam) cultures draw snakes encompassing most of IAU Sco but with fewer head stars. In addition, there are smaller figures: a Mayan Rattlesnake (adjacent to only one Sco star), and a Bororo Turtle drawn over the head of IAU Sco.
	\item[fish] Two Polynesian (Kiribati, Manus) and two Austronesian (Bugis, Mandar) cultures see the IAU Sco region similarly: a ray fish inhabits most of the Northern Sco stars, and the Southern tail stars form the fin of a shark. (In two of these cultures, the shark shape is a naturalistic figure including other stars outside IAU Sco). These two fish constellations form a scene: to the Manus people, the shark bites the stingray. 
\end{description}

\begin{table}[htb]
\caption{\small {\bf Semantic parallels per sky region.} Per pair {\bf IAU region}--{\bf semantic} category, phylogenies with semantic parallels are marked. The exact semantics are also summarised. Root stars with (a) 2+ constellations per phylogeny, and (b) 10+ constellations in the global data are included. Metrics shown: the {\bf similarity} score (rounded), and the (\#{\bf stars} $\ge 2$) in common (when 3+ phylogenies have parallels, these are ranges). The total \#{\bf constellations} is the sum across phylogenies.} 
\label{tab:S1_results_overview}
\centering
{\scriptsize
{\renewcommand{\arraystretch}{0.95} 

	\begin{tabular}{| c ll rrr | p{0mm}p{1mm}p{0mm}p{0mm}p{0mm}p{0mm}p{0mm}p{2mm} | }
		\hline
        & & & & & & \multicolumn{8}{c|}{\bf phylogenies with parallels} \\
		\rot{\bf IAU region} & {\bf semantic} category & exact {\bf semantic} & \rot{\bf similarity} & \rot{\#{\bf stars}} & \rot{total \#{\bf constell.}} & 
			\rot{China} & \rot{Mesopotamia} & \rot{N America} & \rot{S America} & \rot{Polynesia} & \rot{Austronesia} & \rot{Austroasia} & \rot{Main. SE Asia} \\
		\hline

\rowcolor{fill}
{\bf Aqr} & humanoid & both genders & 43\;\; & 2 & 11 & \ch & \me &  &  &  &  &  &  \\
 & group & group of animals & 5\;\; & 2 & 4 &  & \me &  &  &  &  & \as &  \\
\multirow{-2}*{\bf Aur} & humanoid & male worker or myth character & 22\;\; & 4 & 11 &  & \me & \na &  &  &  &  &  \\
\rowcolor{fill}
 & group & council & 26\;\; & 2 & 5 & \ch &  & \na &  &  &  &  &  \\
\rowcolor{fill}
\multirow{-2}*{\bf Boo}& humanoid & male worker & 8+ & 4 & 18 & \ch & \me & \na &  &  &  &  &  \\
{\bf Cap} & mammal & cow, goat & 302\;\; & 2 & 14 & \ch & \me &  &  &  &  &  &  \\
\rowcolor{fill}
 & humanoid & typically female & 6\;\; & 5 & 11 &  & \me & \na &  &  &  &  &  \\
\rowcolor{fill}
\multirow{-2}*{\bf Cas} & man-made object & diverse objects & 2\;\; & 5 & 7 &  & \me & \na &  &  &  &  &  \\
 & group & two humans, dogs, pointers & 2+ & 2 & 10 &  & \me &  & \sa & \po &  & \as &  \\
 & humanoid & both genders & 9+ & 2-4 & 12 & \ch & \me &  &  &  & \an &  &  \\
\multirow{-3}*{\bf Cen} & man-made object & diverse objects & 5\;\; & 2 & 6 &  &  & \na &  & \po &  &  &  \\
\rowcolor{fill}
{\bf Cep} & man-made object & arc-shaped object & 3\;\; & 3 & 5 & \ch & \me &  &  &  &  &  &  \\
{\bf CrA} & man-made object & circle-shaped object & 8+ & 5-7 & 9 &  & \me &  & \sa & \po &  &  &  \\
\rowcolor{fill}
 & group & circular council or dance & 4\;\; & 7 & 4 &  & \me & \na &  &  &  &  &  \\
\rowcolor{fill}
\multirow{-2}*{\bf CrB} & man-made object & circle-shaped object & 2+ & 7 & 15 & \ch & \me & \na &  & \po &  &  &  \\
 & fish & triggerfish, stingray & 5\;\; & 4 & 6 &  &  &  &  & \po & \an &  &  \\
\multirow{-2}*{\bf Cru} & geometric & cross & 54+ & 4 & 10 &  & \me & \na &  & \po &  &  &  \\
\rowcolor{fill}
{\bf Del} & man-made object & arc-shaped object & 5+ & 4-5 & 9 & \ch &  & \na &  & \po &  &  &  \\
{\bf Eri} & landscape & large garden or river & 454\;\; & 14 & 14 & \ch & \me &  &  &  &  &  &  \\
\rowcolor{fill}
 & group & two related entities & 9\;\; & 2 & 7 &  & \me & \na &  &  &  &  &  \\
\rowcolor{fill}
 & humanoid & male pair or myth character & 12\;\; & 5 & 12 &  & \me & \na &  &  &  &  &  \\
\rowcolor{fill}
\multirow{-3}*{\bf Gem} & man-made object & typically a container & 1\;\; & 2 & 4 &  &  & \na &  &  &  &  & \ms \\
{\bf Her} & humanoid & typically male & 15\;\; & 6 & 10 &  & \me & \na &  &  &  &  &  \\
\rowcolor{fill}
 & humanoid & male myth character, dead body & 35+ & 3-7 & 11 & \ch &  & \na &  &  &  & \as &  \\
\rowcolor{fill}
 & mammal & lion, horse, rat & 69\;\; & 6 & 9 &  & \me &  &  & \po &  &  &  \\
\rowcolor{fill}
\multirow{-3}*{\bf Leo} & man-made object & arc-shaped object & 1+ & 6 & 4 &  & \me & \na &  & \po &  &  &  \\
{\bf Lyn} & humanoid & male myth character & 38\;\; & 2 & 7 & \ch &  & \na &  &  &  &  &  \\
\rowcolor{fill}
 & body part & typically arm, leg, hand & 5\;\; & 4 & 6 &  &  & \na & \sa &  &  &  &  \\
\rowcolor{fill}
 & group & three humans, animals, objects & 1+ & 3-10 & 32 & \ch & \me & \na & \sa & \po & \an & \as &  \\
\rowcolor{fill}
 & humanoid & male myth character & 4+ & 7-12 & 15 & \ch & \me & \na & \sa &  &  &  &  \\
\rowcolor{fill}
 & landscape & path or enclosure & 6\;\; & 3 & 5 &  &  &  & \sa & \po &  &  &  \\
\rowcolor{fill}
\multirow{-5}*{\bf Ori} & man-made object & plough, fire, line-shaped objects & 1+ & 3-10 & 36 &  & \me & \na & \sa & \po & \an & \as &  \\
 & arthropod & scorpion & 115+ & 17-18 & 14 &  & \me & \na &  &  &  & \as &  \\
 & fish & shark, stingray & 13\;\; & 13 & 7 &  &  &  &  & \po & \an &  &  \\
\multirow{-3}*{\bf Sco} & reptile & snake, rarely turtle & 18+ & 11-13 & 7 &  &  & \na & \sa &  &  & \as &  \\
\rowcolor{fill}
 & mammal & diverse four-legged animal & 5+ & 5-8 & 14 &  & \me & \na & \sa &  &  &  &  \\
\rowcolor{fill}
\multirow{-2}*{\bf Sgr} & man-made object & dipper, basket, pot & 1\;\; & 8 & 9 & \ch & \me &  &  &  &  &  &  \\
 & body part & animal jaws & 18\;\; & 5 & 6 &  & \me &  & \sa &  &  &  &  \\
\multirow{-2}*{\bf Tau} & man-made object & tongs, tweezers, net & 6\;\; & 6 & 10 & \ch &  &  &  & \po &  &  &  \\
\rowcolor{fill}
 & group & four or seven humans or animals & 4+ & 7 & 19 &  & \me & \na &  & \po &  & \as &  \\
\rowcolor{fill}
 & mammal & bear, elk, caribou, fisher & 5\;\; & 11 & 15 &  & \me & \na &  &  &  &  &  \\
\rowcolor{fill}
\multirow{-3}*{\bf UMa} & man-made object & dipper, boat, cart, net & 1+ & 7 & 29 & \ch & \me & \na & \sa & \po & \an & \as &  \\
 & abstract & compass point & 39\;\; & 2 & 5 & \ch &  &  &  & \po &  &  &  \\
\multirow{-2}*{\bf UMi} & man-made object & dipper, cart, net  & 4\;\; & 7 & 9 &  & \me & \na &  &  &  &  &  \\
		\hline
	\end{tabular}
}
}
\end{table}

\begin{figure}[htb]
	\centering

	\includegraphics[scale=0.6]{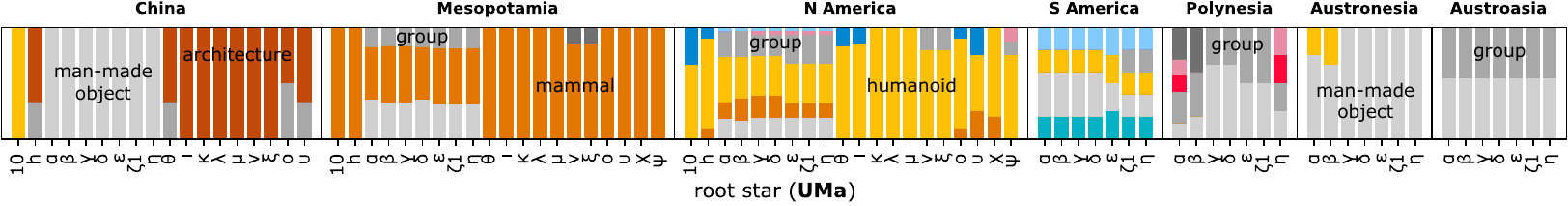}

	\caption{{\small{\bf Semantic parallels for IAU UMa.} Legend as for Figure~\ref{fig:breakdown_semantics_Sco} (top).}}
	\label{fig:breakdown_semantics_UMa} 
\end{figure}

\begin{figure}[htb]
	\centering

	\includegraphics[scale=0.6]{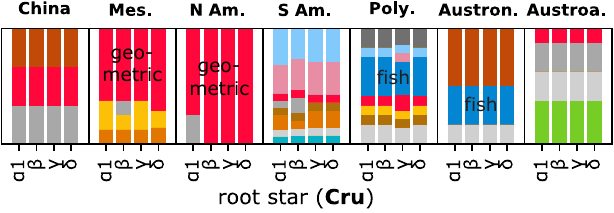}\hspace{5mm}\hspace{10mm}\includegraphics[scale=0.6]{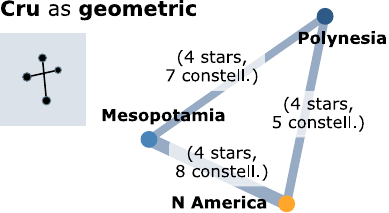}\hspace{10mm}\includegraphics[scale=0.6]{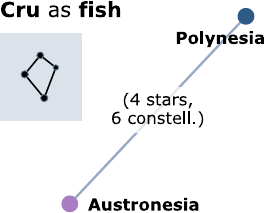}

	\caption{{\small{\bf Semantic parallels for IAU Cru.} Legend as for Figure~\ref{fig:breakdown_semantics_Sco}.}}
	\label{fig:breakdown_semantics_Cru}
\end{figure}

\begin{figure}[htb]

	\includegraphics[scale=0.6]{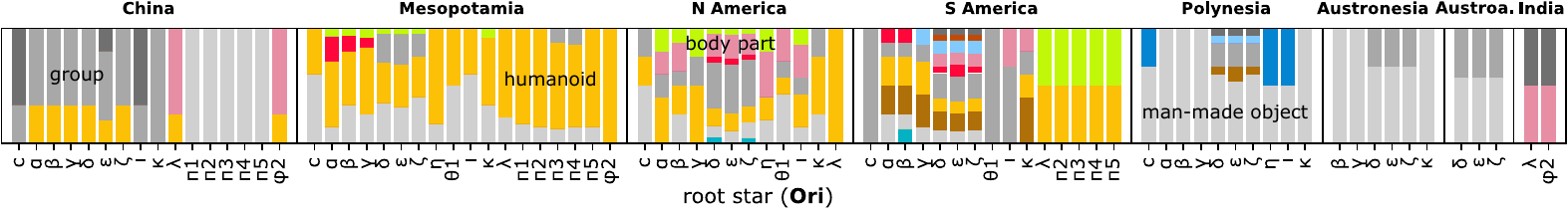}\\[1mm]
	\includegraphics[scale=0.6]{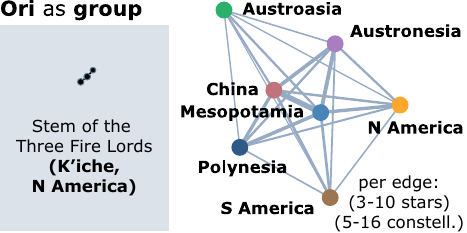}\hspace{10mm}\includegraphics[scale=0.6]{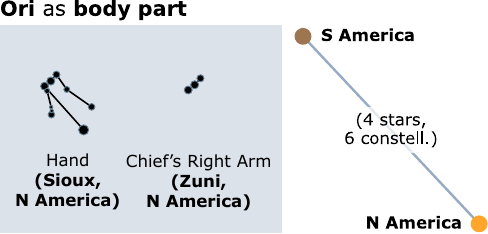}\hspace{10mm}\includegraphics[scale=0.6]{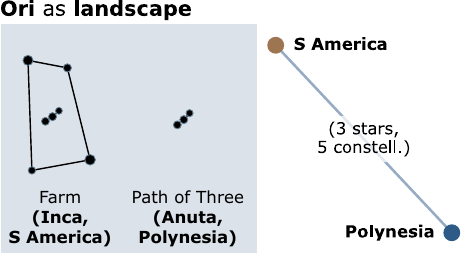}\\[1mm]
	\includegraphics[scale=0.6]{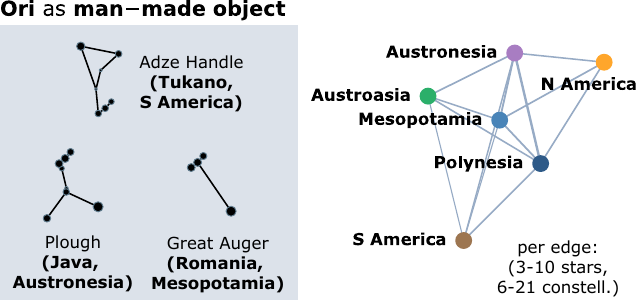}\hspace{10mm}\includegraphics[scale=0.6]{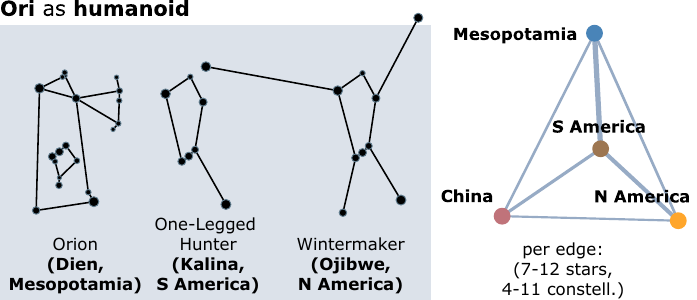}

	\caption{{\small{\bf Semantic parallels for IAU Ori.} Legend as for Figure~\ref{fig:breakdown_semantics_Sco}.}}
	\label{fig:breakdown_semantics_Ori}
\end{figure}

{\bf Summary of results.} Table~\ref{tab:S1_results_overview} summarises all significant semantic parallels per IAU sky region, over groups of 2+ stars. Few of these parallels were measured and explained before: the scorpion in IAU Sco~\cite{kelley2005exploring}, and the many symbolisms of Big Dipper and Orion's belt~\cite{gibbon1964asiatic,gibbon1972asiatic,berezkin2005cosmic,berezkin2012seven}. We discuss the parallels below by their semantic category. Note that further parallels exist for 9 single stars; e.g., $\alpha$ CMi alone is part of spatially large and very bright geometric figures (see Supplementary Table~1 for single-star parallels). The breakdown in semantics is also shown for IAU UMa (Figure~\ref{fig:breakdown_semantics_UMa}), Cru (Figure~\ref{fig:breakdown_semantics_Cru}), and Ori (Figure~\ref{fig:breakdown_semantics_Ori}).

\textbf{Group}. There is prior evidence that some parallels in the group semantic of a few sky regions are cultural (i.e., supported by common myths around IAU UMa, Ori in pre-Columbian N America and central and W Asia~\cite{gibbon1964asiatic,gibbon1972asiatic,berezkin2005cosmic,berezkin2012seven}). However, group constellations occur not only in these regions but worldwide (see for example the breakdown in semantics for UMa in Figure~\ref{fig:breakdown_semantics_UMa}). Due to this, and since their exact group semantic varies widely, the \emph{natural effect} likely dominates. (This statement will gain more weight in the study of semantic parallels by star pattern, below in Sec.~\ref{sec:S2}.) The group symbolism for the belt of {\bf IAU Ori} recurs worldwide, \emph{far broader} than the common hunting myth in N America and Asia, in which the three stars are animals being hunted~\cite{gibbon1972asiatic}. A group of three is present in almost all phylogenies, though diverse in makeup and only rarely related to hunting: fishermen (Norse, Manus), threshers (Russia), runners (Inuit), sons (Sami), saintly women (Romania, Sardinia, Mocoví), deer (Pawnee, Pardhi, Tewa), mountain sheep (Maricopa, Seri), birds (Ava Guaraní), stars, moons, or fire lords (China, Gond, Mandar, K'iche', Samoa), as well as unnamed entities (in other cultures). The seven stars in a near circle in {\bf IAU CrB} are a group of dancing women (Romania, Sardinia) or a council of men (Pawnee, Maricopa). In the Big Dipper ({\bf IAU UMa}), the theme of a stretcher (borne by four men) or bed, followed by a train of mourners or thieves~\cite{gibbon1964asiatic} occurs isolatedly in Central India, but much more often we find Big Dipper as a group of seven: thieves (Macedonia), brothers (Sardinia, Blackfoot), Buddhas (Mongolia), caribou (Inuit), or unnamed entities (Hawaii, Zuni). The group symbolism for Big Dipper does not dominate in any cultural region (Figure~\ref{fig:breakdown_semantics_UMa}), but is broadly present. 

\textbf{Body part}. The Western asterism V of Taurus in {\bf IAU Tau} has cross-phylogeny parallels not as an entire mammal, but as a mammal's jaws. The species varies with the location: a bull (in Babylonia), a wolf (for the Norse in northern Europe), a tapir (for the Lokono and Tupi in S America), or a cayman (for the Tikuna in S America). The simple geometry and distant locations point to a \emph{natural effect} of the jaw symbolism. In N and S America, the {\bf IAU Ori} belt stars sometimes form the basis (joint) of a hand, arm, or leg constellation (for the Sioux, Zuni, Lokono, Tikuna), with or without additional stars. Figure~\ref{fig:breakdown_semantics_Ori} (bottom) shows two examples. This recurrence may be due to \emph{cultural} influence between these neighbouring continents.

\textbf{Man-made object}. The objects represented have simple geometry. The constellation lines can be called naturalistic in all cases, which points to a strong \emph{natural effect}: the star patterns simply resemble these objects. We split the objects in three categories, by shape, as follows.

\emph{Angular objects.} Related to the jaw symbolism (see the body-part semantic), {\bf IAU Tau} is also common as an object similar in function and V shape to jaws, i.e., to hold or capture items: a net (on Manus Island and around China), tongs, tweezers, or perch (on Anuta, Samoa, Vanuatu, and Tonga in Polynesia). In {\bf IAU Ori}, the plough (and related tools: axe, adze, auger, rake, pole, yoke, stick) is widespread, reflecting a common agricultural theme in many regions, but also the natural T shape of Orion's belt and sword (three examples in Figure~\ref{fig:breakdown_semantics_Ori}). Other implements (canoes, strings, traps, nets, fires and other tools) are each present in a few cultures (e.g., the Maya and Aztec marked two fire places in this region). The five stars of {\bf IAU Cas} recur as diverse utilitarian objects: plough or cutter (Babylonia, Macedonia, Seri); lamp stand or container (Inuit); chair or throne (Romania and Sardinia). These figures are composed of 3-5 of the IAU Cas stars, occasionally adding neighbouring stars. 

\emph{Dipper objects.} The Big and Little Dippers ({\bf IAU UMa, UMi}) recur as objects (for Big Dipper, worldwide). Some are objects for handling liquids: a dipper or handful (in the West, around China, and for the K'iche' in N America), and fishing nets, canoes, boats, or ships (for four Austronesian cultures, Marshall in Polynesia, Maricopa in N America, Kalina in S America). Others are carts or beds (in the West and Central India). The cart, bed, and dipper semantics were already assumed \emph{natural}~\cite{gibbon1964asiatic}. On the other hand, the boat constellations are lined in diverse ways (using stars beyond the Big Dipper), so can rather be ascribed to a \emph{cultural} influence or theme in the oceanic ancestries. In {\bf IAU Sgr}, the Western asterisms of the Milk Dipper and Teapot have parallels around China, where dippers and baskets are drawn around the same stars. 

\emph{Arc-shaped objects.} The five stars of {\bf IAU CrA} ($\alpha$ to $\delta$ and $\theta$) are a circle-shaped object in the West, three Polynesian, and two S-American cultures: crown or garland (in the West and the Marshall Islands), fishing net or hook (Manus, Kiribati), wheel (Bororo), leather hide (Mapuche). The seven stars of {\bf IAU CrB} ($\alpha$ to $\epsilon$, $\iota$, $\theta$) are seen similarly: as a crown or garland (in the West and the Marshall Islands), coiled thong (China), net (Carolines, Iroquois), bowl (Tewa), and round table (Macedonia). {\bf IAU Del} also recurs as arc-shaped objects: a bowl, gourd, or trough of the 3-4 northern stars (in China, Carolines, Kiribati, and the Marshall Islands); or a longer tool of 5 stars: sling, bow (Pawnee, Tewa, Zuni), or adze (Anuta). The three brightest stars of {\bf IAU Cep} ($\alpha$ to $\gamma$) recur both in the West and around China as part of arc-shaped objects: a hook, cover, arc, or scythe. The hook of {\bf IAU Leo} is a sickle or adze (in the West, and Samoa), a fish trap (Kiribati), fireplace (Sioux). All objects are arcs; an additional Tewa constellation of only three stars is a travois (an A-shaped frame).

\textbf{Geometric}. The cross symbolism for the four bright stars ($\alpha$ to $\delta$) of {\bf IAU Cru} is likely a recent, Western \emph{cultural diffusion}. It recurs across six phylogenies (Figure~\ref{fig:breakdown_semantics_Cru}, left), but in only three (Mesopotamian, N American, and Polynesian), the cross occurs more than once, so we call this a parallel (Figure~\ref{fig:breakdown_semantics_Cru}, right). Nine cultures draw the same four-star non-planar cross: Western cultures, three present-day Mesoamerican (Huave, K'iche', Tzotzil), and one 20th-century Polynesian tradition (Manus Island). Additionally, this cross is also listed in post-18th-century China (known to have hosted European Jesuit astronomers, who developed charts for the southern celestial pole in the 17th century~\cite{hashimoto2004jesuit}), but not in medieval China nor Korea (where these four stars are integrated in more complex asterisms with different meanings). Only in Hawaii, Cru breaks the pattern: $\gamma$ Cru alone is the end star in a line of stars sky-wide from the north to the south pole, a metaphor for a genealogical line.

\textbf{Fish, reptile, arthropod}. The same four {\bf IAU Cru} stars recur instead with a fish semantic in six cultures of Polynesian and Austronesian ancestry (Caroline Islands, Kiribati, Marshall Islands, Samoa, Madura, and Malay)---usually a triggerfish, but a stingray for the Malay. It is usually lined into a diamond (Figure~\ref{fig:breakdown_semantics_Cru}, right). Interestingly, the parallel of the fish symbolism in {\bf IAU Sco} is between the same two ancestries (Figure~\ref{fig:breakdown_semantics_Sco}, bottom) and also around specific fish species---ray fish and shark. Both may be a \emph{cultural diffusion}, along common fishing themes in these neighbouring regions. Surprisingly, an expected fish symbolism, in {\bf IAU Del}, is rare outside Western cultures: a single Polynesian constellation over these stars is a fish (the Kailou Fish of the Manus). On the other hand, cultural contamination alone is an unlikely cause for the recurrence, in {\bf IAU Sco}, of the scorpion, from the present tribes of Central India to pre-colonial Mesoamerica and ancient Mesopotamian traditions, as well as the reptile, between Central India and tribes in both Americas. These are instead likely a \emph{natural effect} of the star pattern in IAU Sco, which was also shown to have near-universal line geometry regardless of semantics~\cite{bucur2022pone}.

\textbf{Humanoid}. Many sky regions depict humanoids, but the composition of the star group, as well as the line geometry, vary wildly. Thus, this recurrence can only be ascribed to a \emph{natural effect}: the star patterns are so complex that human figures of \emph{any size} and \emph{any degree of naturalism} can be drawn creatively. For {\bf IAU Aur, Boo, Gem}, the Western constellations are naturalistic humanoids, some N-American constellations incorporate the bright stars of these regions into much larger, whole-sky naturalistic male figures, while Chinese asterisms only draw small and abstract lines. The five stars in {\bf IAU Cas} ($\alpha$ to $\epsilon$) form a female in four Western cultures, as well as for the Navajo and Yellowknives of N America; other N-American cultures (Ahtna, Lower Tanana) add many other stars to create whole-sky male figures (of which Cas forms one hand). {\bf IAU Leo} also has no universal line figure, nor size of the star group: the Leo ``hook'' is part of a large emperor figure in China and Korea, but is a hunter figure for the Huave in Mesoamerica; the five stars in the lion's torso form a naturalistic dead body in a funeral procession for the Gond in Central India; various Leo stars are also part of whole-sky humanoid figures in five N-American cultures. {\bf IAU Ori} is a humanoid across four phylogenies (diverse in shape but always naturalistic, except in China); examples in Figure~\ref{fig:breakdown_semantics_Ori} (bottom). Surprisingly, this semantic occurs in both hemispheres and in the tropics, so even when the orientation of the star group changes, a human figure is still sometimes seen in this sky region.

\textbf{Mammal}. The mammal symbolism rarely crosses phylogenetic borders: only four IAU sky regions have this parallel. Also here the star groups and line designs vary, so the recurrence can be ascribed to a \emph{natural effect} in most cases. In {\bf IAU Cap}, an ox or cow is drawn in the China region; this uses only two bright stars ($\alpha^2, \beta^1$ Cap) into a detailed line figure, much smaller than the Western horned goat. {\bf IAU Sgr} (a human-mammal hybrid in Western cultures) is a mammal in two N- and two S-American cultures, but the species (fox, deer, guanaco, feline) and line figures are diverse, because the star pattern in this region is complex. In non-Western cultures where {\bf IAU UMa} is not the typical large bear over many stars, but mostly limited to the Big Dipper, the mammal  varies (an elk in Russia; a bear, fisher, caribou, and a dog mentioned by the Tewa, all in N America) but the same tail likely made the mammal symbolism appear natural. Surprisingly, an expected mammal symbolism, in {\bf IAU Tau}, is rare outside Western cultures: there is a single S-American constellation (the Tapir of the Kalina, identical in star makeup to the V of Taurus, so likely representing only the tapir's jaws---see the paragraph below on parallels in body-part symbolism). Besides the known \emph{cultural diffusion} of the bear semantic for the Big Dipper in the West and N America~\cite{gibbon1964asiatic}, we find only one other instance: {\bf IAU Leo} recurs as a mammal in Polynesia, as a Hawaiian lion of likely modern borrowing from the West. 

\textbf{Landscape}. {\bf IAU Eri}, a long chain of faint stars representing a long river in eight Western cultures, surprisingly recurs as other elements of landscape. In China and Korea, halves of the same star chain form two landscape elements: in the southern region between $\upsilon^1$ and $\iota$, a hill or orchard; in the northern region between $\gamma$ and $\tau^9$, a garden or meadow. All line figures are curved chains of stars. In Eri, constellations are not drawn in other cultures (except in Polynesia, and then only in the northern region). Orion's belt in {\bf IAU Ori} recurs as a path (for the Tupi of S America, and the Polynesian cultures on Tonga and Anuta Island); also, the bright rectangle of Orion enclosing the belt is a farm or garden in S America. For both sky regions, the parallels can be assumed to be a \emph{natural effect} of the star pattern.

These results show that there are numerous sky regions in which symbolism appears to follow the shape and star makeup of the constellation. While for humanoids and mammals the shape of the star group frees the human eye to see complex and creative figures (rather than guiding the eye into a fixed shape), for other semantic categories (group, body part, man-made object, reptile, and arthropod) the association is clear-cut: symbolism matches shape across cultural regions. The natural effect of the sky regions onto symbolism can thus be called universal in the sky regions and cultural regions marked in Table~\ref{tab:S1_results_overview}. In S.2 below, we pose a stronger question.

\subsection{Question (S.2): Semantic universality per star pattern}
\label{sec:S2}

We ask \emph{to what extent does this natural effect generalise}: not only to the peculiarities of a fixed sky region (such as the reptile recurring in IAU Sco in the Americas and Austroasia, per Table~\ref{tab:S1_results_overview}), but sky-wide to similar-looking star patterns. When semantic parallels across cultural regions are found regardless of the sky region where they occur, there is stronger evidence for the \emph{natural effect} upon constellation symbolism.

To study the association between star pattern and symbolism, we first quantify the star pattern into numerical features (see Section~\ref{sec:s2method} for details). After feature selection (which eliminates correlated ones), eight remain. They describe the appearance of the star group from different points of view, and are expected to affect constellation symbolism:

\begin{itemize}
	\setlength\itemsep{0em}
	\item the {\bf number of stars} in the star group;
	\item the {\bf aspect ratio} of the star group's point pattern;
	\item its {\bf spatial diameter}, in degrees on the celestial sphere;
	\item the {\bf fraction of stars on the convex hull}, which shows the internal complexity of the pattern (circular and linear patterns both have values close to 1, while random patterns have low values); the {\bf fraction of stars in line} on the spatial Minimum Spanning Tree (MST) drawn over the star group (a purely linear MST will have value 1); relatedly, the {\bf average MST branching} is higher the more branched the MST is;
	\item the {\bf minimum} and {\bf maximum magnitude} among the stars.
\end{itemize}

The number of stars, the spatial diameter, and the magnitude features are used in their absolute values; the remaining features are normalised to $[0,1]$ (dividing by the size of the star group), such that they are size-independent.

We then train machine-learning classifiers to \emph{describe the statistical association} between the features of the star group and the semantic category of that constellation. The classifiers are Support-Vector Classifiers (SVC) with balanced class weights and regularisation (more details in Section~\ref{sec:s2method}). If a classifier can discriminate a semantic category well, then an association exists between star pattern and that semantic. The association is then \emph{interpreted} via classification maps: plots showing which feature values associate with which semantic. The strength of a statistical pattern is seen in the following performance metrics (all normalised to $[0,1]$). The balanced \emph{accuracy} is the fraction of constellations whose semantic category was correctly discriminated (with the categories weighed equally). This accuracy should be judged against its \emph{random baseline}: 1 divided by the number of categories. The \emph{recall} and \emph{precision} are per category. Usually a semantic category has high recall but low precision: the interpretation is that the semantic does have a characteristic star pattern, but other semantics share that same star pattern. Low precision is unavoidable, since no star group is assigned a single semantic worldwide (as seen before in Figure~\ref{fig:breakdown_semantics_phylogeny}, bottom).

Worldwide, the classifier finds a weak association between star pattern and semantic (accuracy 0.37, baseline 0.07) for the 14 semantic categories represented in the global data (see Supplementary Figure~1 for the confusion matrix). Few semantics have good recall worldwide (plant: 0.67, geometric: 0.58, arthropod: 0.56, reptile: 0.55). Stratified by phylogeny, stronger associations are found, and they differ among cultural regions. (Supplementary Figure~1 shows all confusion matrices.) China and Mesopotamia (accuracies 0.55-0.60, baseline 0.09) have different semantics which match strongly to a specific star pattern (in terms of recall, Chinese bird: 1.00, plant: 0.91, body part and mammal: 0.90; Mesopotamian landscape: 0.92, arthropod and reptile: 0.81). Semantics can't be distinguished well in N and S America as well as Polynesia (accuracies 0.44-0.50, baselines 0.09-0.13); in recall, the best are the S American and Polynesian body parts: 0.69 vs.\ 0.83. Precision is variable and generally low, because a diversity of semantics is assigned to the same stars.

To interpret why certain semantics are assigned to certain star patterns, we isolate each phylogeny-semantic combination and train a focused, binary classifier (accuracy baseline 0.50) to discriminate that semantic from all others in that phylogeny. This is a simpler problem which requires less data, so can be done for the smallest phylogenies, such as Egypt. A classifier may find an association between semantic and star pattern among the cultures of that phylogeny. It is the next step which matters though: if the same association is found across multiple phylogenies (like for question S.1), only then we call that a semantic parallel. If there is no known cultural transmission cross-phylogeny, this is likely a \emph{natural effect}: astronomical traditions from multiple phylogenies interpret star patterns similarly.

{\bf Summary of results.} Six semantics are assigned to the same type of star pattern across phylogenies (summary in Table~\ref{tab:S2_results_overview}), all clearly \emph{natural effects}. We discuss them below.

\begin{table}[!htb]
\caption{\small {\bf Semantic parallels per star pattern.} For each pair of {\bf type of star pattern} and {\bf semantic}, phylogenies with semantic parallels are marked. Only semantics with 5+ constellations, and phylogenies with 2+ such semantics, are included. Performance metrics for binary classifiers are shown as ranges across phylogenies.}
\label{tab:S2_results_overview}
\centering
{\scriptsize
{\renewcommand{\arraystretch}{1.1} 

	\begin{tabular}{| p{6.2cm} p{1.4cm} rrr | p{0mm}p{1mm}p{0mm}p{0mm}p{0mm}p{0mm}p{0mm}p{0mm}p{2mm} | }
		\hline
        & & & & & \multicolumn{9}{c|}{\bf phylogenies with parallels} \\
		type of {\bf star pattern} & {\bf semantic} & accuracy & recall & precision & 
			\rot{China} & \rot{Mesopotamia\;} & \rot{Egypt} & \rot{N America} & \rot{S America} & \rot{Polynesia} & \rot{Austronesia} & \rot{Austroasia} & \rot{India}\\
		\hline
		\rowcolor{fill}
		relatively few stars (2-8), small spatial diameter (under 26 deg.), on linear MST and on convex hull & abstract & .71-.80 & .67-.90 & .27-.72 &  &  &  & \na &  & \po &  &  & \india \\
		like abstract, but additionally: stars with similar magnitude & geometric & .82-.89 & .91-.95 & .19-.25 &  & \me &  & \na &  &  &  &  &  \\
		\rowcolor{fill}
		relatively few stars (2-8), small spatial diameter (under 26 deg.), on linear MST & group (1) & .72-.91 & .78-1 & .20-.52 &  & \me &  & \na & \sa & \po &  & \as & \\
		\rowcolor{fill}
		$\qquad$additionally: stars are bright, with similar magnitude & group (2) & .76-.91 & .78-.97 & .20-.50 &  & \me &  &  & \sa & \po &  &  & \\
		aspect ratio and MST branching are low and correlate & arthropod & .88-.93 & .88-1 & .21-.26 &  & \me &  & \na &  &  &  &  &  \\
		\rowcolor{fill}
		low aspect ratio \emph{or} low MST branching & reptile (linear) & .73-.84 & .83-1 & .06-.17 & \ch & \me &  & \na & \sa &  &  &  & \\
		\rowcolor{fill}
		high aspect ratio \emph{and} high MST branching & reptile (spiral) & .73-.84 & .83-1 & .06-.17 & \ch & \me &  & \na & \sa &  &  &  & \\
		many stars, large spatial diameter, non-linear MST, stars rarely on convex hull & humanoid (naturalistic) & .73-.98 & .71-1 & .33-.83 &  & \me & \eg & \na &  &  &  &  & \\
		few stars (2-6), small spatial diameter (under 20 deg.) & humanoid (abstract) & .71-.72 & .85-1 & .08-.23 & \ch & \me &  & \na &  & \po &  &  &\\
		\hline
	\end{tabular}
}
}
\end{table}

\begin{figure}[htb]
	\centering

	\includegraphics[scale=0.5]{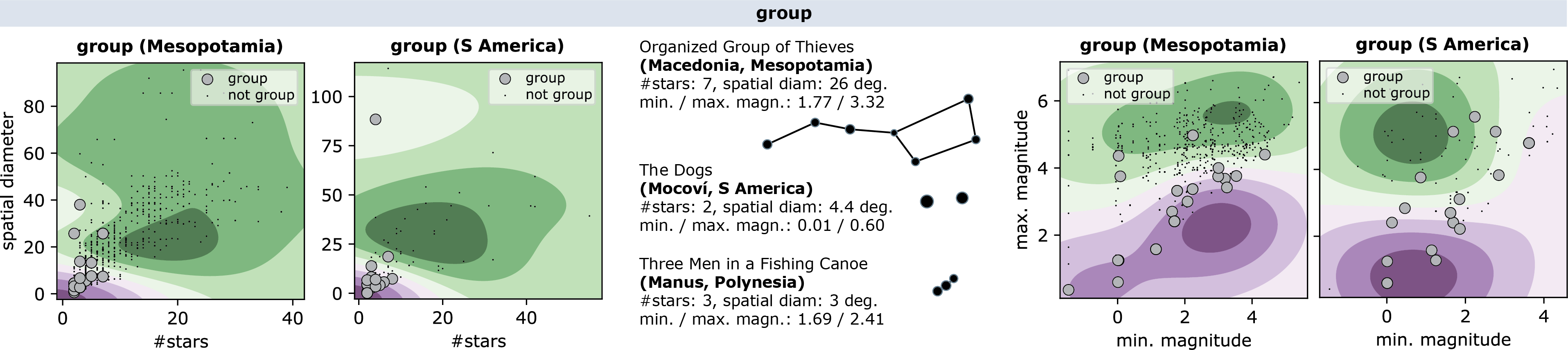}\\[1mm]
	\caption{{\small{\bf Semantic parallels per star pattern (group).} Probabilistic, binary classification maps show the decision boundaries between this semantic and all others, by features of the star pattern. {\bf\textcolor{Purple}{Purple}} areas denote the one semantic of interest, and {\bf\textcolor{Green}{green}} areas all others. Example constellations are also shown.}}
	\label{fig:classification_maps_group}
\end{figure}

\textbf{Group.} Groups are distinguished in five cultural regions: group constellations are typically star patterns of 2-8 stars with a small spatial diameter (under 26 deg., the diameter of the Big Dipper), and with a linear MST. Figure~\ref{fig:classification_maps_group} shows classification maps for pairs of important features, and examples. The pair min.-max.\ magnitude shows a distinct property of group constellations in only three cultural regions (the West, S America, and Polynesia): the stars are bright and comparable in magnitude. There are a few exceptions: groups with a dominant bright star and a trail of faint stars, such as the Romanian constellation She-Goat with Three Kids (identical to IAU Lyr). 

\textbf{Geometric and abstract.} Constellations named after lines (straight or bent), closed polygons, crosses, letters of the Latin alphabet with simple geometries (G, W, Y, V), zigzags, or circles are well represented only in Western and N American cultures. In India and Polynesia, we find abstract symbolisms instead: the names are adjectives rather than nouns (e.g., blessed, red, gifted, bountiful, famous). The star patterns are similar except in magnitude. The star patterns of constellations called after a geometric shape have relatively few stars (2-8) similar in brightness, typically all lying on a purely linear MST (fr.\ MST in line and fr.\ on convex hull are usually maximum, 1), and on the convex hull (a feature which can differentiate them from group constellations). The aspect ratio varies, but the spatial diameter is mostly small (below 26 deg., with a few exceptions in Western cultures: the largest are the Western asterisms Winter Hexagon and Heavenly G, with 66 deg.).

\begin{figure}[htb]
	\centering

	\includegraphics[scale=0.5]{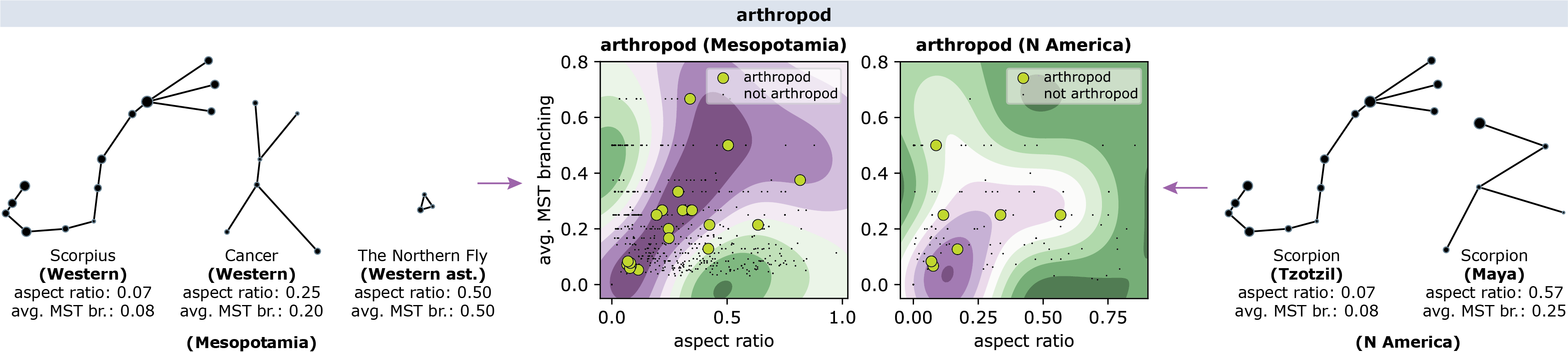}\\[1mm]
	\includegraphics[scale=0.5]{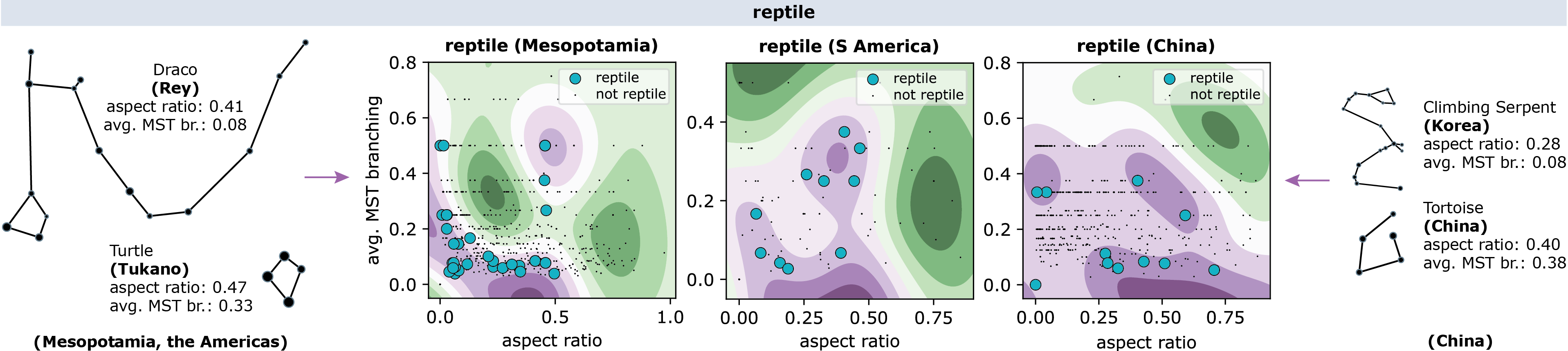}\\[1mm]
	\caption{{\small{\bf Semantic parallels per star pattern (arthropod and reptile).} Legend as for Figure~\ref{fig:classification_maps_group}.}}
	\label{fig:classification_maps_artrep}
\end{figure}

\textbf{Arthropod and reptile.} Only in Western cultures and in N America, arthropods are drawn over star patterns which are relatively elongated (aspect ratio mostly below 0.5, but not 0), and relatively branched rather than linear (avg. MST branching mostly below 0.5, but not 0). Figure~\ref{fig:classification_maps_artrep} shows classification maps, as well as examples. The typical arthropod is in the lower left corner of the map, and is a scorpion (or, rarely, a spider or crab). Outliers towards higher aspect ratios and MST branching are other species (crabs or flies). Most scorpions are similar to IAU Sco in star makeup and line figure, so this semantic parallel owes a lot to the high frequency of arthropods drawn in the sky region of IAU Sco by these cultures (Table~\ref{tab:S1_results_overview}). 
For reptiles, like for arthropods, the aspect ratio and MST branching of the star pattern are the most important discriminants (Figure~\ref{fig:classification_maps_artrep}), and the association is consistent across four phylogenies. However, the star pattern needed for reptiles is different: the branching of the MST for reptile must be low. There are two types of reptile constellations. The dominating type has either linear (low aspect ratio) or coiled MST (high aspect ratio)---unbranched. This allows to draw snakes (either straight or coiled) by largely following the MST. A less frequent type has both high aspect ratio and high MST branching: these are usually spiral snakes or turtles with near-circular shapes. Drawing reptiles thus appears universal, and can be called naturalistic in all cases.

\begin{figure}[htb]
	\centering

	\includegraphics[scale=0.5]{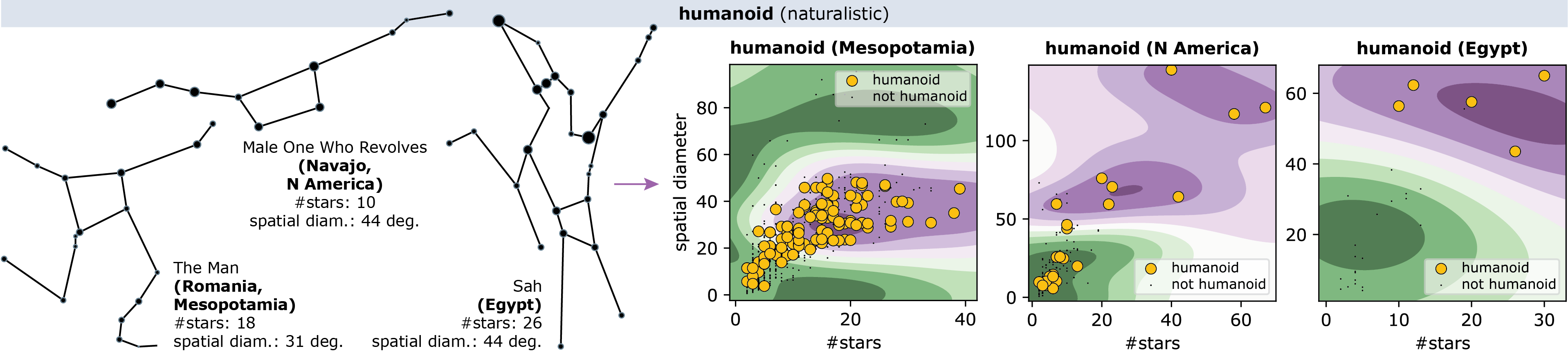}
	\caption{{\small{\bf Semantic parallels per star pattern (humanoid).} Legend as for Figure~\ref{fig:classification_maps_group}. Supplementary Figure 2 shows more discriminative features: the fraction of MST in line, and the fraction on the convex hull.}}
	\label{fig:classification_maps_hum}
\end{figure}

\textbf{Humanoid.} The number of stars and spatial diameter of the group are important discriminants in Western, Egyptian, and N-American cultural regions: many stars in spatially large groups are made into humanoids (Figure~\ref{fig:classification_maps_hum}). These star patterns have complex inner structure: the MST is non-linear, and the stars are rarely on the convex hull (Supplementary Figure 2). There are exceptions: a fraction of small humanoid constellations break the pattern in Western and N-American cultures, and instead match Chinese and Polynesia humanoids, for which the association is the opposite: with few exceptions, the smallest groups with few stars are made into humanoids. These are simple chains or coils of 2-6 stars with variable brightness, fundamentally different from the complex line figures in Western cultures. There are thus not one, but \emph{two} ways to imagine humans in the sky: either naturalistically onto expansive star groups of many stars, or abstractly onto few stars.

\section{Discussion}

\textbf{Novelty, limitations, and future work.} This study provides a systematic, worldwide measurement of parallels in the meaning assigned to constellations by traditional astronomies. This is different from the studies which quantified cross-culture parallels in either the geometric structure of the constellations figures~\cite{bucur2022pone}, or the star group~\cite{kemp2020perceptual}---without considering constellation symbolism. It is also unlike prior work which only showed examples of semantic parallels in few sky regions (IAU UMa, Ori, plus Pleiades which is out of scope here) for N America and Eurasia~\cite{gibbon1964asiatic,gibbon1972asiatic,berezkin2005cosmic,berezkin2012seven}. 

We give an alternate point of view: instead of the depth of being able to explain an instance of cultural effect which diffused the symbolism of one asterism on some continents~\cite{berezkin2005cosmic,berezkin2012seven}, we provide geographic breadth, a sky-wide analysis, and means to quantify semantic parallels and their statistical significance---all for more general insight. Many surprising findings are of note per sky region (Table~\ref{tab:S1_results_overview}): from the widespread presence of man-made objects among the semantics of many sky regions to the landscape semantic for the faint sky region of IAU Eri. Future work may provide depth to these results, for example by finding new evidence to explain semantic parallels by cultural diffusion, for some of the sky regions where diffusion is a possible cause. Future work may also quantify the extent to which the geometry of the line figure is itself predictable from the star group alone.

Our results do come with limitations: they hold for the current dataset of constellations. Precise data on astronomical traditions is scarce, and folk traditions are fading fast; further data, if available (particularly in the least well covered geographic regions) may help uncover more semantic parallels. Our choice of semantic categories can be considered subjective: some semantic categories may be too broad. For example, snakes look different than turtles, although both are reptiles---we have observed the two categories as two types of star patterns in the results of Question S.2, but may have missed this difference in other semantic categories. The largest semantic category (man-made objects) could also be split into subcategories, which would mean that stronger requirements would be placed on the type of object that is matched between cultures. Despite the limitations, we found many previously unknown facts about constellation design, which we summarise as follows.

\textbf{Semantic parallels have a geographic scope.} There is evidence of a natural effect: the shape of the star group does associate with (and likely affected, as per our causal model in Figure~\ref{fig:causal_model}) the constellation semantics. This is seen at a glance from the fact that many of the parallels in Tables~\ref{tab:S1_results_overview}-\ref{tab:S2_results_overview} expand broadly across cultural regions, to an extent unexplainable by cultural diffusion. Semantics which have a characteristic star pattern (Table~\ref{tab:S2_results_overview}) can be called \emph{endogenous} (or \emph{predictable}) to that type of star pattern, from the point of view of humans. Likewise but less generally, semantics which are characteristic to a sky region (Table~\ref{tab:S1_results_overview}) can be called endogenous to that sky region. For both research questions (S.1-2), any association is geographically bounded to certain phylogenies, and---since few semantic categories are well represented worldwide---the semantic parallels themselves are rarely worldwide. For example, there is a neat geographic demarcation of the three semantics for IAU Sco (Figure~\ref{fig:breakdown_semantics_Sco}): the sky region is named an arthropod, fish, or reptile in almost entirely distinct geographic regions, even though reptiles in general live worldwide. Also, from the results, it becomes apparent that, despite not being a large phylogeny, N America (8.5\% of the constellations in the data, from largely oral traditions) has so much diversity among its cultures that it registers semantic parallels with other cultural regions as frequently as Western cultures (34.7\% of the constellations, from both literate and oral traditions). 

\textbf{There are principles behind the symbolism of constellations.} These findings are complementary to the known parallels in line geometry~\cite{bucur2022pone} and star group~\cite{kemp2020perceptual}. The semantic parallels discovered here point to a natural effect, and are evidence of some \emph{universality of semantic constellation design} (Table~\ref{tab:S2_results_overview}). This universal design matches intuition (e.g., drawing naturalistic humanoids would require a star cluster with many stars and a non-linear MST, groups would fit a star cluster with few stars and a linear MST, and geometric shapes would need similar stars placed on the convex hull). We \emph{quantify} these newly found constellation design `principles' by measuring and summarising the features of characteristic star patterns. 

These principles recall and agree with known Gestalt principles of human perception. Both geometric and group constellations are perceived when a linear MST is composed of few stars. Reptiles and arthropods are also perceived over MSTs with low branching, so almost linear. Abstract and geometric constellations are formed over stars which lie on their convex hull. The MST of a star group minimises the total length of the lines, thus implementing the \emph{Gestalt principle of minimality}; when the shape of the MST is a determining factor, this principle is supported. When the MST is also linear, it supports the \emph{Gestalt principle of continuity} (lines prefer to form wide angles). When stars lie on the convex hull, this recalls the \emph{Gestalt principle of closure}~\cite{zahn1971graphgestalt,wagemans2012century}; all are global metrics over a point pattern, and aid the perception of clusters. The results thus support the view that natural perception shapes not only the structure, but also the symbolism of constellations, and this is to an extent valid cross-culture.



\section{Data and method}
\label{sec:data_and_method}

\subsection{Constellation data}
\label{subsec:data}

The dataset consists of constellations from astronomical cultures worldwide. A minority were contributed to the astronomy software Stellarium~\cite{Stellarium} by members of the public; we use this data after validating it against existing scholarly sources, and verifying licenses. The majority of the cultures were digitised by the author from scholarly sources; some supplement existing Stellarium cultures. The data is public at \url{https://github.com/doinab/constellation-lines} and is a living dataset. It was partially introduced in a prior analysis on the network signature of constellation line figures~\cite{bucur2022pone}. \textbf{Supplementary Table~2} overviews the data and the many data sources.

\textbf{Inclusion criteria and limitations.} Each culture contains a number of asterisms or constellations (here, all typically called constellations). We use all documented constellations for which the stars were identified; in almost all cases, there is also a line figure. For oral cultures, the data was documented with variable degrees of certainty from sources with imprecise text or sky charts. Single-star asterisms, tight star clusters such as the Pleiades, and other sky objects such as galaxies, are excluded. See our prior study \cite{bucur2022pone} (the Data section) for a detailed description of the \emph{inclusion criteria}, \emph{limitations} of the data collection (approximate star identification, unknown culture size due to lost oral knowledge, a potential bias towards recalling bright stars), \emph{types} of cultures (astronomical literacy, practical use for the constellations), and a regional \emph{timeline} of constellation records.

\textbf{Location and phylogeny.} The {\bf phylogeny} is a \emph{region of cultural influence or migration}. In summary:

\begin{description}
	\item[Mesopotamia] The Western zodiac originates in Babylonia (as early as \textasciitilde3200 BC), where it represented gods and associated animals. These were borrowed by the Greeks (\textasciitilde500 BC) and transmitted to the West~\cite{rogers1998origins1}. We mark all astronomical cultures with Western influence (such as the standard set from the International Astronomical Union, IAU) as having Mesopotamian ancestry. This naming, however, is an \emph{oversimplification}; Western cultures are a mix of the two traditions: additional constellations of probable Mediterranean origin were also assembled into the Greek tradition~\cite{rogers1998origins2}, and gaps in the sky were also filled with minor constellations in modern times. We use this name for the phylogeny to show the oldest root origin, even though some of the constellations in this set have different, more recent origins. This phylogeny contains most European folk astronomies, including now obsolete attempts, such as the earliest modern, lined star charts by French astronomers Ruelle and Dien~\cite{Ruelle,Dien} (dated 1786 and 1831). The only exception is the {\bf Sami} culture, which has no known continental influence, so forms a phylogeny of its own.
	\item[Egypt] is a single, ancient culture, from which we use only the older, native-Egyptian constellations (combined with the Mesopotamian in the later classical era~\cite{rogers1998origins1}). These were reconstructed from pictographic sources: the astronomical ceiling of the tomb of Senenmut at Deir el Bahari in Luxor (\textasciitilde1470 BC), and the Egyptian figures on the Dendera zodiac~\cite{lull2009constellations}. 
	\item[India] The Indian moon stations (27-28 in number, including single stars which are not used here) were documented during the Vedic period (before 500 BC). No external influence in this early period is known~\cite{basham1954wonder}.
	\item[China] Ancient Chinese astronomy (developed before 1000 BC) is literate, so well documented. It likely developed without external influences until the 17th c. and its influence spread to Korea and Japan. 
	\item[N America] groups regional cultures from the Arctic to the south west, the Great Plains~\cite{miller1997stars} and Mesoamerica. With the exception of the reconstructed Aztec and Maya astronomies, they are recently documented.
	\item[S America] groups cultures located on both coasts and in the Amazon. Since some tribes have migrated across this continent~\cite{magana1984tupi}), they are grouped into a single region of cultural influence. They are also relatively recently documented, with the exception of the reconstructed Inca and Tupi traditions.
	\item[Polynesia] groups Polynesian and culturally Polynesian islands, with ancestry in a common seafaring culture.
	\item[Austronesia] groups cultures on or around the islands of Sulawesi and Java (Indonesia), and Peninsula Malaysia. They (including the Austronesian Malay subgroup) are connected by an Austronesian migratory tradition and were recently documented. Only the Thai culture is marked as having a separate phylogeny ({\bf Mainland SE Asia}), due to unclear or mixed influence: it is documented as indigenous~\cite{orchiston2021seasia} and empirically shows commonalities to the Malay asterisms, despite known early influence from China and India.
	\item[Austroasia] groups populations from the Austroasiatic family, a migration which is distinct from the Austronesian. These are tribes from Central India (a mix of genetically Austroasians and ancestral Indo-Europeans) and the Nicobarese tribe of Nicobar Islands (speaking a language from the Austroasiatic family). The level of contamination or modification by interaction with other tribes seems to be low~\cite{vahia2013gond}.
\end{description}

See \cite{bucur2022pone} (the Data section) for the reasoning behind the phylogeny annotations. These are not always clear due to lack of historical evidence; we use the best knowledge available.

\subsection{Semantic annotation}
\label{subsec:sem}

We annotate by hand the semantic of each constellation in the dataset. This is a judgement done by the \emph{meaning} (the main object named), not the \emph{shape} of the constellation (for example, the Chinese asterism Eunuch is not figurative in shape, but semantically it is a humanoid). Rarely, there are exceptions which pose difficulties: a constellation has \emph{alternative} names (in which case, we mark the two best alternative semantics), has \emph{mixed} symbolism (Sagittarius is a centaur, half human and half horse, drawing a bow, so is marked as both humanoid and mammal), is \emph{ambiguous} (we make a choice: the Korean constellation Four Spirit of River is marked as a landscape, not a group), or is \emph{unnamed} or unintelligible (in which case, it is not annotated with any semantic).

Note that the same constellation has occasionally \emph{changed semantics} in time and across cultures. We mark semantics separately by culture, so are able to capture these changes. The Greek Capricorn (a goat) was formerly the Mesopotamian Goat-Fish (which has a fish lower half, and a goat upper half, so is assigned to both fish and mammal semantics). Ten Mesopotamian constellations have preserved the same stars but gained different names upon borrowing by the Greeks: the former Agricultural Worker (a humanoid) became Aries (a mammal, renamed by a scribe's mistake), and the former Swallow (a fish and a swallow bird, touching tails) became Pisces (two fish)~\cite{pettinato1998scrittura}.

The semantic categories follow biological taxonomy (for constellations naming living organisms), or otherwise a taxonomy denoting the type of object. We settled on the following semantic categories:

\begin{description}
	\item[humanoid] (usually one, rarely two) individual human(s), each represented with some detail, on 2+ stars (regardless of line geometry);
	\item[animal] by biological group (usually one, rarely two, each represented on multiple stars): \textbf{mammal}, \textbf{reptile} (including amphibian), \textbf{bird}, \textbf{fish} (including mollusc), \textbf{arthropod} (usually scorpion, crab, insect, spider);
	\item[body part] may be a head, leg, arm, jaw, back, tentacle, eyes, claws, hand, hair, etc.;
	\item[group] two or more entities of any type, each represented without detail, so stars represent individuals; also when part of the constellation is a group, and part an object, it is marked as group;
	\item[geometric] may be a (bent, zigzag) line, triangle, cross, quadrilateral, hexagon, letters (G, W, Y, V), or circle;
	\item[man-made object] a mobile object: a tool, grill, table, chair, boat, adze, cart, plough, bowl, spear, net, dipper, basket, bed, tomb, cup, flag, fire place, stove, etc.;
	\item[architecture] an immobile man-made construction: a house, gate, room, office, kitchen, steps, passageway, tower, lodge, wall, fence, well, pillars, poles, bridges, doors, etc.;
	\item[landscape] outdoors scenery: a river, mountain, garden, farm, field, territory, hill, yard, enclosure, encampment, market, orchard, cloud, rain, thunder, lightning, sea, river, pool, street, road, path, etc.;
	\item[abstract] a concept without a shape (e.g., an emptiness, force, death, bond, voice, amount), or an under-described, non-humanoid entity (a demon, ogre, a spirit), or a geographic pointer (northern pole, west, mark, star).
\end{description}

In particular, the semantic categories of {\bf man-made object}, {\bf architecture}, and {\bf landscape} can be considered subjective. Some elements of architecture are figuratively similar to some elements of landscape (a wall is like a path), but both categories are sufficiently numerous (9.2\% of the constellations denote architecture, and 5.5\% landscape) to stand on their own. Man-made objects are numerous (22.6\% of the constellations), but were not split into subcategories because many objects of different scales but similar function look similar, and the notion of scale does not apply in the sky (a fish hook is like a fishing net or a garland; a boat is like a dipper or a bowl).

\subsection{Computational analyses for Question (S.1): Semantic universality per sky region}
\label{sec:s1method}

Semantic universality in a sky region is said to be present if the semantic \emph{similarity score} between pairs of phylogenies is high. Given a semantic $s$ and two phylogenies $i$ and $j$, denote by $f^s_i$ and $f^s_j$ the fractions of constellations from each phylogeny which are from that semantic. The joint probability of that semantic between these phylogenies is $f^s_i \cdot f^s_j$. 

Then, take a root star $r$ which is represented in constellations from both phylogenies. As criteria for representation, we ask that a root star has at least (a) 2 constellations per phylogeny, and (b) 10 constellations in the global data. Among the constellations incident on $r$, the same fractions from above are instead denoted $f^s_i(r)$ and $f^s_j(r)$, and the joint probability of having that semantic for constellations incident on $r$ is $f^s_i(r) \cdot f^s_j(r)$. Then, the semantic similarity score averages this among all root stars $r$ in common between the two phylogenies:
\[
\text{similarity score}^s_{ij} = \dfrac{\text{average}_r \; f^s_i(r) \cdot f^s_j(r)}{f^s_i \cdot f^s_j}
\]

A similarity score is positive, but unbounded. Scores below 1 are not considered (they mean that similarity exists, but is expected by the natural semantic makeup of the phylogenies). High scores can be considered significant.

Two additional counts are reported: the \emph{number of root stars} in that sky region, and the \emph{number of constellations} in common for that semantic between the two phylogenies. There are caveats to interpreting these numbers. If there are $n$ root stars in common, it is possible that one phylogeny consistently draws a constellation such that it uses all $n$ stars, but the other splits the $n$ stars into two groups and consistently draws two smaller constellations, all with the same semantic. Also, it is likely that different cultures from a single phylogeny sometimes use different subsets of the $n$ stars to draw constellations with that semantic; the union of these subsets is taken per phylogeny, and the number of root stars in common between two phylogenies is reported as the intersection of two such unions. This allows internal variation for how the sky region is grouped into constellations of the semantic of interest, so does not require perfect matches to be made (which would produce very few results). 

\subsection{Computational analyses for Question (S.2): Semantic universality per star pattern}
\label{sec:s2method}

The {\bf star-pattern features} were already listed in Section~\ref{sec:S2}. We describe how the more complex ones are computed. 

The {\bf aspect ratio} of the star group's point pattern is estimated by first translating the star coordinates from polar to 3D Cartesian, then computing the eigenvalues of the covariance matrix for the point cloud (functions for this are available in linear-algebra libraries, such as {\tt numpy}). Intuitively, an eigenvalue is the factor by which a characteristic vector (a direction) is stretched. The aspect ratio is the ratio of the second largest to the largest eigenvalue, and ranges in $[0, 1]$. 

For the remaining features, the convex hull and the spatial Minimum Spanning Tree (MST) are first calculated. Both are subsets of the Delaunay triangulation~\cite{berg1997computational} on a sphere over the point cloud, which is computed with the {\tt stripy} triangulation package~\cite{Moresi2019}. The MST is the ``backbone'' subset of this triangulation, such that all stars are linked into a line figure, and this line figure has minimal global length (i.e., sum of line lengths on the celestial sphere). The MST is computed with the {\tt networkx} package. The convex hull is a subset of the triangulation, such that only the edges which are part of exactly one triangle are kept. Some stars of the constellation star group may lie inside this hull. 

The \textbf{fraction of stars on the convex hull} is the ratio of stars on the convex hull, out of all stars in the group. The \textbf{fraction of stars in line on the MST} is the normalised hop diameter of the MST (the fraction of stars on the MST's longest branch). The \textbf{average MST branching} takes only the stars which are not MST ``leaves'' (are ``inner'' nodes, i.e., have a degree strictly higher than 1). The feature is the sum of their degree (each degree minus 1, to discount the node's parent in the tree) divided by the number of stars (minus 1, to discard the root of the tree). A low value corresponds to a many-star, linear MST. A high value (which cannot reach exactly 1) corresponds to a branched MST. A caveat to interpreting these values is that a line of three can be considered both linear and branched (its average MST branching value is 0.5). 

{\bf Machine-learning classifiers} are used here only to \emph{describe statistical patterns}, not to predict the semantic of a new constellation (in other words, there is no test data). The classifiers are multi-class Support-Vector Classifiers (SVC) implemented in the package {\tt scikit-learn}~\cite{scikit-learn}. They are configured to train with a strength of the regularisation $C$ depending on the number of semantic categories to classify (between 2 and 14) and the amount of data. (The strength of regularisation is inversely proportional to $C$.) $C=1$ is sufficient regularisation for a binary classifier, and $C=100$ for the much harder problem of multi-class classification. We ascertain that the model neither over- nor underfits by inspecting the classification maps, which should not build decisions around single constellations.

The \emph{accuracy} of discrimination among semantic categories is the fraction of constellations whose semantics were correctly discriminated (\emph{balanced} such that all semantic categories weigh equally). The \emph{baseline} (1 divided by the number of semantic categories) would be achieved by a random classifier, or, alternatively, by a classifier which always predicts one class. The \emph{recall} is the fraction of constellations from that semantic which were correctly predicted. The \emph{precision} is the fraction of constellations predicted to have a semantic which are indeed from that semantic. We configure the classifiers to train with balanced class weights, which tends to lead to low precision but high recall; this is to make the classifiers learn equally about the smaller class (the constellations from the semantic of interest) as about the dominant class (the constellations \emph{not} from the semantic of interest).



{\footnotesize
	\bibliographystyle{plain}
	\bibliography{main}
}

\end{document}


\maketitle

\section*{Results}

\subsection*{Question (S.1): Semantic universality per sky region}

Supplementary Table~\ref{tab:S1_results_overview_supp} summarises all significant semantic parallels per IAU sky region, for constellations which overlap over single stars only. For example, $\alpha$ CMi alone is part of spatially large and very bright geometric figures (Western asterisms depicting: an X-shaped Egyptian Cross, the Winter Triangle, the Winter Hexagon, and the spiral of a letter G; as well as the Great Cross of the Quechua and the Celestial Hexagon of the Tukano in S America). Since a single star is in common, the geometry of the constellations is necessarily diverse, and only the semantic categories have parallels.

\begin{table}[!htb]
\caption{\small {\bf Semantic parallels per sky region.} Per pair {\bf IAU region}--{\bf semantic} category, phylogenies with semantic parallels are marked. The exact semantics are also summarised. Root stars with (a) 2+ constellations per phylogeny, and (b) 10+ constellations in the global data are included. Metrics shown: the {\bf similarity} score, rounded. The total \#{\bf constellations} is the sum across phylogenies.} 
\label{tab:S1_results_overview_supp}
\centering
{\scriptsize
{\renewcommand{\arraystretch}{0.95} 

	\begin{tabular}{| c ll rrr | p{0mm}p{1mm}p{0mm}p{0mm}p{0mm}p{0mm}p{0mm}p{2mm} | }
		\hline
        & & & & & & \multicolumn{8}{c|}{\bf phylogenies with parallels} \\
		\rot{\bf IAU region} & {\bf semantic} category & exact {\bf semantic} & \rot{\bf similarity} & \rot{\#{\bf stars}} & \rot{total \#{\bf constell.}} & 
			\rot{China} & \rot{Mesopotamia} & \rot{N America} & \rot{S America} & \rot{Polynesia} & \rot{Austronesia} & \rot{Austroasia} & \rot{Main. SE Asia} \\
		\hline
	\rowcolor{fill}
	$\gamma$ {\bf And} & humanoid & various humans & 28\;\; & 1 & 10 & \ch & \me &  &  &  &  &  &  \\
	$\alpha$ {\bf CMi} & geometric & triangle, cross, hexagon, one letter G & 175\;\; & 1 & 6 &  & \me &  & \sa &  &  &  &  \\
	\rowcolor{fill}
	$\zeta$ {\bf Cep} & humanoid & male characters & 31\;\; & 1 & 8 & \ch & \me &  &  &  &  &  &  \\
	$\sigma$ {\bf Lib} & man-made object & scales, chariot & 11\;\; & 1 & 9 & \ch & \me &  &  &  &  &  &  \\
	\rowcolor{fill}
	$\beta$ {\bf Oph} & humanoid & male myth characters, bureaucrats, gods & 34\;\; & 1 & 10 & \ch & \me &  &  &  &  &  &  \\
	$\upsilon$ {\bf Per} & humanoid & various humans & 29\;\; & 1 & 7 & \ch & \me &  &  &  &  &  &  \\
	\rowcolor{fill}
	$\beta$ {\bf Tau} & humanoid & male myth characters & 6\;\; & 1 & 6 &  & \me & \na &  &  &  &  &  \\
	10 {\bf UMa} & humanoid & male myth characters & 34\;\; & 1 & 5 & \ch &  & \na &  &  &  &  &  \\
	\rowcolor{fill}
	$\beta$ {\bf UMi} & man-made object & dipper, cart, yoke & 5\;\; & 1 & 10 &  & \me & \na &  &  & \an &  &  \\
		\hline
	\end{tabular}
}
}
\end{table}

\subsection*{Question (S.2): Semantic universality per star pattern}

\begin{figure}[htb]
	\centering

	\includegraphics[scale=0.5]{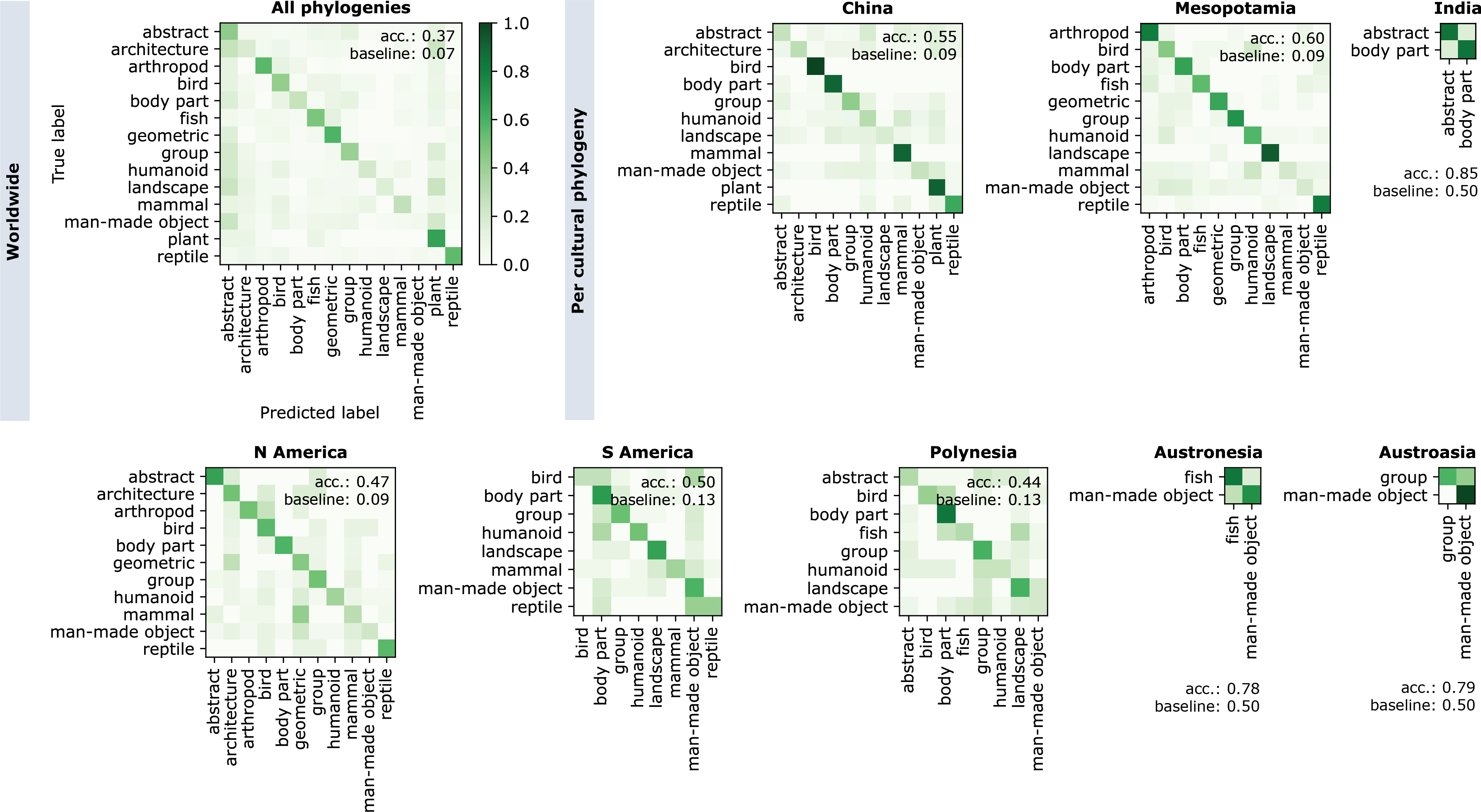}
	\caption{{\small{\bf Semantic universality by star pattern.} (Worldwide, then per phylogeny:) Normalised confusion matrix for semantics, given the features of the star pattern. Only semantics with at least 5 constellations, and phylogenies with 2 or more such semantics, are included.}}
	\label{fig:CM-allsem}
\end{figure}

Stratified by phylogeny (and only looking at semantics represented in each phylogeny by 5+ constellations), some statistical associations between star pattern and semantic are found by multi-class classifiers. The strength of these associations vary among phylogenies. For an overview, Supplementary Figure~\ref{fig:CM-allsem} shows confusion matrices: each cell represents the fraction of constellations from a semantic which are correctly predicted to be from that semantic. A diagonal matrix would mean a perfect, one-to-one association between a star pattern and a semantic.

\begin{figure}[htb]
	\centering

	\includegraphics[scale=0.5]{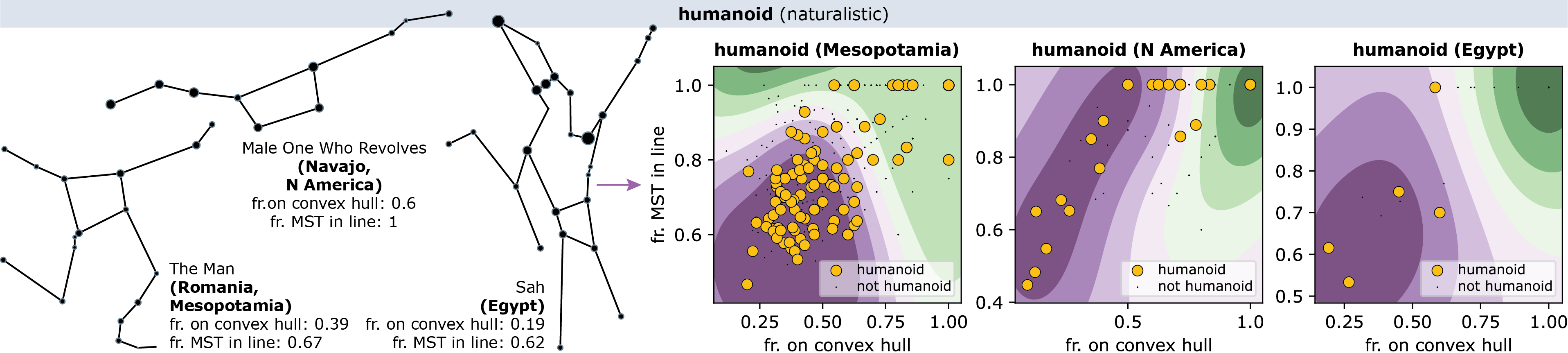}
	\caption{{\small{\bf Semantic parallels per star pattern.} Probabilistic, binary classification maps show the decision boundaries between one semantic and all others, by features of the star pattern. {\bf\textcolor{Purple}{Purple}} areas denote the one semantic of interest, and {\bf\textcolor{Green}{green}} areas all others. Example constellations are also shown.}}
	\label{fig:classification_maps_supp}
\end{figure}

\section*{Data}

26 small cultures were added to the dataset since the publication of \cite{bucur2022pone}, as well as a small number of constellations for which only the stars (but not the lines) could be identified. All constellations have at least 2 stars. Single stars, star clusters with small diameters (such as the Pleiades), and other sky objects (such as visible galaxies) are out of scope for this study. Supplementary Table~\ref{tab:culture_list} below presents an overview of the dataset.

{\footnotesize
{\renewcommand{\arraystretch}{0.96} 
\begin{longtable}[H]{ | rcl lcrr | }
\caption{\footnotesize {\bf Sky cultures.} Phylogeny is marked: M (Mesopotamia), E (Egypt), In (India), S (Sami), nA and sA (North and South America), C (China), P (Polynesia), An (Austronesia), Aa (Austroasia), mseA (Mainland SE Asia).\label{tab:culture_list}} \\
	\hline
	$\quad\;\;${\bf location}	&	{\bf phylogeny} & {\bf sky culture}	& {\bf timestamp}	& {\bf source}	& {\bf \#constell.}	& {\bf references} \\
	\hline
	\endfirsthead 
	\hline
	$\quad\;\;${\bf location}	&	{\bf phylogeny} & {\bf sky culture}	& {\bf timestamp}	& {\bf source}	& {\bf \#constell.}	& {\bf references} \\
	\hline
	\endhead
	\rowcolor{fill}
	Global		& M	& \textbf{IAU}			& 105 AD--20th c.& standard		& 88	& \cite{IAU, Stellarium} \\
	\rowcolor{fill}
	Global		& M	& \textbf{Rey}			& 1952			& book			& 80	& \cite{rey1952stars, Stellarium} \\
	\rowcolor{fill}
	Global		& M	& \textbf{Western}		& present		& dataset		& 88	& \cite{Stellarium} \\
	\rowcolor{fill}
	Global		& M	& \textbf{Western asterisms} & present	& dataset		& 53	& \cite{Stellarium} \\
	N Africa	& E	& \textbf{Egypt}		& 1470,  50 BC 	& carving, paper & 26 	& \cite{lull2009constellations, Stellarium} \\
	\rowcolor{fill}
	W Asia		& M	& \textbf{Babylonia}	& 1100-700 BC	& tablet, papers& 50	& \cite{hoffmann2017hipparchs, hoffmann2017history, hoffmann2018drawing, Stellarium} \\
	\rowcolor{fill}
	W Asia		& M	& \textbf{Al-Sufi}		& 964 AD		& book, dataset	& 51	& \cite{Al_Sufi_Book, Stellarium} \\
	\rowcolor{fill}
	W Asia		& In& \textbf{Arabia moon st.}	& 9th c.	& book, paper	& 21	& \cite{kunitzsch2013lunar, Stellarium} \\
	S Asia		& Aa& \textbf{Banjara}		& present		& paper			& 1		& \cite{vahia2014banjaraskolam} \\
	S Asia		& Aa& \textbf{Gond}			& present		& paper			& 11	& \cite{vahia2013gond} \\
	S Asia		& Aa& \textbf{Kolam}		& present		& paper			& 7		& \cite{vahia2014banjaraskolam} \\
	S Asia		& Aa& \textbf{Korku}		& present		& paper			& 7		& \cite{vahia2016korku} \\
	S Asia		& Aa& \textbf{Nicobars}		& present		& paper			& 3		& \cite{vahia2018nicobars} \\
	S Asia		& Aa& \textbf{Pardhi}		& present		& paper			& 6		& \cite{halkare2019pardhi} \\
	S Asia		& An& \textbf{Bugis}		& present		& papers		& 12	& \cite{orchiston2021mandar, orchiston2021seasia} \\
	S Asia		& An& \textbf{Java}			& 19th c.		& book, paper	& 3		& \cite{van1885astronomie, khairuddin2021austronesian} \\
	S Asia		& An& \textbf{Madura}		& present		& paper			& 6		& \cite{fatima2021madura} \\
	S Asia		& An& \textbf{Malay}		& present		& paper			& 3		& \cite{jaafar2021folk} \\
	S Asia		& An& \textbf{Mandar}		& present		& paper			& 6		& \cite{orchiston2021mandar} \\
	S Asia		& In& \textbf{India moon st.}& < 500 BC		& book, dataset	& 21	& \cite{basham1954wonder, Stellarium} \\
	S Asia		& mseA& \textbf{Thai}		& present		& paper			& 9		& \cite{orchiston2021seasia} \\
	\rowcolor{fill}
	E Asia		& C	& \textbf{China medieval}& 1092 AD		& chart, book	& 245	& \cite{pan1989history, Stellarium} \\
	\rowcolor{fill}
	E Asia		& C	& \textbf{China}		& 1756-1950		& chart, book	& 252	& \cite{pan1989history, Stellarium} \\
	\rowcolor{fill}
	E Asia		& C	& \textbf{Japan moon st.} & 8th c. 		& chart, paper	& 27	& \cite{renshaw2000cultural, Stellarium} \\
	\rowcolor{fill}
	E Asia		& C	& \textbf{Korea}		& 1395			& chart, dataset& 218	& \cite{Stellarium} \\
	\rowcolor{fill}
	E Asia		& M	& \textbf{Mongolia}		& present		& dataset		& 4		& \cite{Stellarium} \\
	Eurasia		& M	& \textbf{Russia}		& present		& book			& 4		& \cite{svjatskij2007astronomija} \\
	\rowcolor{fill}
	Europe		& M	& \textbf{Belarus}		& 19th-21st c.	& paper			& 14	& \cite{avilin2008astronyms, Stellarium} \\
	\rowcolor{fill}
	Europe		& M	& \textbf{Dien}			& 1831			& chart			& 100	& \cite{Dien} \\
	\rowcolor{fill}
	Europe		& M	& \textbf{Macedonia}	& present		& paper			& 16	& \cite{cenev2008macedonian, Stellarium} \\
	\rowcolor{fill}
	Europe		& M	& \textbf{Norse}		& 13th c.		& verse, book, dataset& 6& \cite{kaalund1908alfraedhi, Stellarium} \\
	\rowcolor{fill}
	Europe		& M	& \textbf{Romania}		& 1907			& book, exhibition& 37	& \cite{otescu1907, crtro, Stellarium} \\
	\rowcolor{fill}
	Europe		& M	& \textbf{Ruelle}		& 1786			& chart			& 74	& \cite{Ruelle} \\
	\rowcolor{fill}
	Europe		& M	& \textbf{Sardinia}		& present		& dataset		& 11	& \cite{Stellarium} \\
	\rowcolor{fill}
	Europe		& S & \textbf{Sami}			& 19th c.		& book			& 4		& \cite{Lundmark1982bmn, Stellarium} \\
	N America	& nA& \textbf{Ahtna}		& present		& thesis		& 2		& \cite{cannon2021northerndene} \\
	N America	& nA& \textbf{Aztec}		& 16th c.		& codices, book	& 4		& \cite{PrimerosMemoriales, FlorentineCodex, aveni1980skywatchers, Stellarium} \\
	N America	& nA& \textbf{Blackfoot}	& 20th c.		& book			& 4		& \cite{miller1997stars} \\
	N America	& nA& \textbf{Gwich'in}		& present		& thesis		& 2		& \cite{cannon2021northerndene} \\
	N America	& nA& \textbf{Huave}		& 1981			& paper			& 15	& \cite{lupo1981conoscenze} \\
	N America	& nA& \textbf{Inuit}		& 20th c.		& book, dataset	& 9		& \cite{macdonald1998arctic, Stellarium} \\
	N America	& nA& \textbf{Iroquois}		& early 20th c.	& book			& 4		& \cite{miller1997stars} \\
	N America	& nA& \textbf{Koyukon}		& 20th c., present& book, thesis& 2		& \cite{miller1997stars,cannon2021northerndene} \\
	N America	& nA& \textbf{Lower Tanana}	& present		& thesis		& 1		& \cite{cannon2021northerndene} \\
	N America	& nA& \textbf{Maricopa}		& 20th c.		& book			& 6		& \cite{miller1997stars} \\
	N America	& nA& \textbf{Maya}			& 15th c.		& codex, books	& 14	& \cite{freidel1993maya, spotak2015maya, Stellarium} \\
	N America	& nA& \textbf{Mi'kmaq}		& late 19th c.	& book			& 4		& \cite{miller1997stars} \\
	N America	& nA& \textbf{Navajo}		& 20th c.		& book			& 6		& \cite{miller1997stars} \\
	N America	& nA& \textbf{Ojibwe}		& present		& book			& 9		& \cite{lee2014ojibwe, Stellarium} \\
	N America	& nA& \textbf{Pawnee}		& 20th c.		& book			& 11	& \cite{miller1997stars} \\
	N America	& nA& \textbf{Sahtúot’įnę}	& present		& thesis		& 1		& \cite{cannon2021northerndene} \\
	N America	& nA& \textbf{Seri}			& present		& book, dataset	& 15	& \cite{blanco2018seri, Stellarium} \\
	N America	& nA& \textbf{Sioux}		& present		& books			& 13	& \cite{lee2014dlakota, miller1997stars, Stellarium} \\
	N America	& nA& \textbf{Tewa}			& 20th c.		& book			& 12	& \cite{miller1997stars} \\
	N America	& nA& \textbf{Tutchone}		& 20th c.		& book			& 1		& \cite{miller1997stars} \\
	N America	& nA& \textbf{Tzotzil}		& 20th c.		& paper, book	& 9		& \cite{vogt1997zinacanteco, milbrath2000star} \\
	N America	& nA& \textbf{Yellowknives}	& present		& thesis		& 3		& \cite{cannon2021northerndene} \\
	N America	& nA& \textbf{Zuni}			& 20th c.		& book			& 9		& \cite{miller1997stars} \\
	C America	& nA& \textbf{K'iche'}		& 20th c.		& paper			& 8		& \cite{tedlock1985hawks} \\
	\rowcolor{fill}
	S America	& sA& \textbf{Ava Guaraní}	& 1921			& paper			& 6		& \cite{lehmann1924chiriguanos} \\
	\rowcolor{fill}
	S America	& sA& \textbf{Bororo}		& 1983			& book			& 16	& \cite{fabian1992space} \\
	\rowcolor{fill}
	S America	& sA& \textbf{Inca}			& 1613			& book			& 8		& \cite{lehmann1928coricancha} \\
	\rowcolor{fill}
	S America	& sA& \textbf{Kalina}		& 1980			& paper			& 10	& \cite{magana1982carib} \\
	\rowcolor{fill}
	S America	& sA& \textbf{Lokono}		& present		& dataset		& 10	& \cite{rybka2018lokono, Stellarium} \\
	\rowcolor{fill}
	S America	& sA& \textbf{Mapuche}		& 19th c.		& book			& 7		& \cite{menares2015mapuche} \\
	\rowcolor{fill}
	S America	& sA& \textbf{Mocoví}		& present		& thesis		& 12	& \cite{lopez2009virgen} \\
	\rowcolor{fill}
	S America	& sA& \textbf{Quechua}		& 20th c.		& paper			& 6		& \cite{urton1978beasts} \\
	\rowcolor{fill}
	S America	& sA& \textbf{Tikuna}		& present		& dataset		& 4		& \cite{Stellarium} \\
	\rowcolor{fill}
	S America	& sA& \textbf{Toba}			& present		& book, papers	& 11	& \cite{lehmann1923tobas1,lehmann1924tobas2,sanchez2010rasgos} \\
	\rowcolor{fill}
	S America	& sA& \textbf{Tukano}		& 1905, 2007	& book, thesis	& 21	& \cite{koch1905anfange, cardoso2007ceu, Stellarium} \\
	\rowcolor{fill}
	S America	& sA& \textbf{Tupi}			& 1614			& book, papers	& 8		& \cite{abbeville1614histoire, magana1984tupi, afonso2013constelaccoes, Stellarium} \\
	\rowcolor{fill}
	S America	& sA& \textbf{Wichi}		& 20th c.		& book, papers	& 5		& \cite{mariani2017look,metraux1939matako,lehmann1923matacos} \\
	Pacific		& P	& \textbf{Anuta}		& 1998			& book			& 11	& \cite{feinberg1988polynesian, Stellarium} \\
	Pacific		& P	& \textbf{Carolines}	& 1951			& paper			& 23	& \cite{goodenough1951native, holton2015east} \\
	Pacific		& P	& \textbf{Hawaii}		& present		& website, dataset& 13	& \cite{HSL, Stellarium} \\
	Pacific		& P	& \textbf{Kiribati}		& present		& dictionary	& 16	& \cite{KED} \\
	Pacific		& P	& \textbf{Manus}		& 20th c.		& paper			& 12	& \cite{hoeppe2000shark} \\
	Pacific		& P	& \textbf{Māori}		& 19th c.		& paper			& 4		& \cite{orchiston2000polynesian, Stellarium} \\
	Pacific		& P	& \textbf{Marshall}		& present		& dictionary	& 41	& \cite{MEOO} \\
	Pacific		& P	& \textbf{Samoa}		& present		& dataset		& 14	& \cite{Stellarium} \\
	Pacific		& P	& \textbf{Tonga}		& late 19th c.	& paper			& 11	& \cite{collocott1922tongan, Stellarium} \\
	Pacific		& P	& \textbf{Vanuatu}		& present		& website, dataset& 6	& \cite{DMR, Stellarium} \\
	\hline
\end{longtable}
}
}

{\footnotesize
	\bibliographystyle{plain}
	\bibliography{main}
}